\documentclass{article}

\usepackage{PRIMEarxiv}

\usepackage{hyperref}       % hyperlinks
\usepackage[utf8]{inputenc} % allow utf-8 input
\usepackage[T1]{fontenc}    % use 8-bit T1 fonts
\usepackage{url}            % simple URL typesetting
\usepackage{booktabs}       % professional-quality tables
\usepackage{amsfonts}       % blackboard math symbols
\usepackage{nicefrac}       % compact symbols for 1/2, etc.
\usepackage{microtype}      % microtypography
\usepackage{lipsum}
\usepackage{fancyhdr}       % header
\usepackage{graphicx}       % graphics
\graphicspath{{media/}}     % organize your images and other figures under media/ folder

\usepackage{todonotes}
\usepackage[acronym, shortcuts, nohypertypes={acronym}]{glossaries}
\newacronym{CBT}{CBT}{cognitive behavioural therapy}
\newacronym{DNN}{DNN}{deep neural network}
\newacronym{EMA}{EMA}{ecological momentary assessment}
\newacronym{LIWC}{LIWC}{\emph{Linguistic Inquiry and Word Count}}
\newacronym{LLM}{LLM}{large language model}
\newacronym{NLP}{NLP}{natural language processing}
\newacronym{RAG}{RAG}{retrieval augmented generation}
\newacronym{SGM}{SGM}{stochastic generative model}

\usepackage[natbib, bibencoding=utf8, citestyle=numeric, bibstyle=ieee, maxbibnames=10, maxcitenames=2, mincitenames=1, sortcites]{biblatex}
\bibliography{references}

\usepackage[capitalize]{cleveref}

\definecolor{orange}{RGB}{255,127,0}
\usepackage[most]{tcolorbox}
\newtcolorbox[auto counter,crefname={Glossary}{boxes}]{pabox}[2][]{colback=orange!25!white,
colframe=orange!75!black,fonttitle=\bfseries,
colbacktitle=orange!85!black,enhanced,
title=Glossary ~\thetcbcounter: #2,
label={#1}}

\title{
Large language models for mental health %--\\In a nutshell
\thanks{\textit{\underline{Citation}}: 
\textbf{Authors. Title. Pages.... DOI:000000/11111.}} 
}

\author{
  Andreas Triantafyllopoulos$^{1,2}$,
  Yannik Terhorst$^{3}$,
  Iosif Tsangko$^{1,2}$,
  Florian B. Pokorny$^{1,2,4}$,\\
  \textbf{Katrin D. Bartl-Pokorny}$^{1,4}$,
  \textbf{Lennart Seizer}$^5$,
  \textbf{Ayal Klein}$^{6,7}$,
  \textbf{Jenny Chim}$^7$,\\
  \textbf{Dana Atzil-Slonim}$^6$,
  \textbf{Maria Liakata}$^7$,
  \textbf{Markus Bühner}$^3$,
  \textbf{Johanna Löchner}$^5$,
  \textbf{Björn Schuller}$^{1,2,8}$\\
  $^1$CHI -- Chair of Health Informatics, Technical University of Munich, Munich, Germany \\
  $^2$MCML -- Munich Center for Machine Learning, Munich, Germany \\
  $^3$Department of Psychology, Ludwig Maximilian University of Munich, Munich, Germany\\
  $^4$Division of Phoniatrics, Medical University of Graz, Graz, Austria\\
  $^5$Friedrich-Alexander University, Erlangen-Nürnberg, Germany\\
  $^6$Department of Psychology, Bar-Ilan University, Ramat Gan, Israel\\
  $^7$Queen Mary University of London, London, UK\\
  $^8$GLAM -- Group on Language, Audio, \& Music, Imperial College, London, UK\\
  \texttt{andreas.triantafyllopoulos@tum.de}
}

\begin{document}
\maketitle

\begin{abstract}
Digital technologies have long been explored as a complement to standard procedure in mental health research and practice, ranging from the management of electronic health records to app-based interventions.
The recent emergence of large language models (LLMs), both proprietary and open-source ones, represents a major new opportunity on that front.
Yet there is still a divide between the community developing LLMs and the one which may benefit from them, thus hindering the beneficial translation of the technology into clinical use.
This divide largely stems from the lack of a common language and understanding regarding the technology's inner workings, capabilities, and risks.
Our narrative review attempts to bridge this gap by providing intuitive explanations behind the basic concepts related to contemporary LLMs.
\end{abstract}

%BS: Always use (at least) 5 keywords if allowed for SEO - I added one :)
\keywords{Digital Psychology \and Clinical Psychology \and Mental Health \and Large Language Models \and Artificial Intelligence}

\section{Introduction}
Mental health disorders are among the leading causes of disease burden worldwide and are projected to be the number one cause of global disability-adjusted life years by 2030~\citep{WorldHealthOrganization2022, Herrman2022, James2018}.
Facing this challenge technological innovations for mental health care are highly needed~\citep{Terhorst2024, Moshe2021, Herrman2022, seiferth2023mental}.
While internet- and mobile-based interventions have been extensively researched in the last three decades~\citep{Buck2024, chang2018electronic}, showing promising results when compared to traditional face-to-face psychotherapy~\cite{Terhorst2024, Moshe2021, hedman2023therapist, cuijpers2019effectiveness}, 
%yt at least this is true, if we ignore the systematic difference in sample characteristics, study design aspects, and conflict of interest situation ;) %yt  this is not true. acceptance levels as well as actual usage are rather low to moderate at best (Philippi, P., Baumeister, H., Apolinário-Hagen, J., Ebert, D. D., Hennemann, S., Kott, L., Lin, J., Messner, E.-M., & Terhorst, Y. (2021). Acceptance towards digital health interventions – Model validation and further development of the Unified Theory of Acceptance and Use of Technology. Internet Interventions, 26, 100459. https://doi.org/10.1016/j.invent.2021.100459) also adherence sucks (Moshe, I., Terhorst, Y., Philippi, P., Domhardt, M., Cuijpers, P., Cristea, I., Pulkki-Råback, L., Baumeister, H., & Sander, L. B. (2021). Digital interventions for the treatment of depression: A meta-analytic review. Psychological Bulletin, 147(8), 749–786. https://doi.org/10.1037/bul0000334) and even worse in real-world (Baumel, A., Edan, S., & Kane, J. M. (2019). Is there a trial bias impacting user engagement with unguided e-mental health interventions? A systematic comparison of published reports and real-world usage of the same programs. Translational Behavioral Medicine, 9(6), 1020–1033. https://doi.org/10.1093/tbm/ibz147) - I know this does not fit the storyline, but the positive remark on acceptance is IMO not true.
user acceptance and adherence has been moderate to low~\citep{Philippi21-ATD, moshe2021digital, baumel2019there}.
To address this, the relatively recent introduction of \acp{LLM} -- models which can process and generate natural language -- and their impressive performance across a variety of \ac{NLP} tasks, has (re-)sparked an overwhelming amount of public and commercial interest in one of the earliest technology-based ideas to augment and supplement mental health practice -- conversational agents~\citep{vaidyam2021changes, abd2019overview, Demszky23-ULL}. 
ELIZA was the first such conversational agent (`chatbot') to be proposed in 1966~\citep{Weizenbaum66-EAC} and was composed out of a set of rules that acted on the input data (text supplied by the human user) to continue a conversation adhering to the basic principles of Rogerian psychology.
Due to its nature, ELIZA was limited to rehashing the text provided by the human user and reformulate it into a question that nudged the user to elaborate on it.
% Despite this simplistic approach, ELIZA has been explored as an intervention against depression -- sometimes with some success.
Following that, studies on chatbots indicated first promising results in reducing depressive symptoms and psychological stress~\citep{Li2023, He23-CAI}, and conversational platforms targeted to mental health, like Woebot\footnote{\url{https://woebothealth.com/}}, are already offered to patients. 

Research -- and computational capabilities -- have progressed substantially since those early days, and modern incarnations like ChatGPT or Claude can have more naturalistic, open-ended conversations.
This has generated a lot of excitement about the potential application of those models in mental health practice.
In particular, the ability to \emph{scale} their deployment to millions of users, coupled with their instantaneous availability, make them one of the most promising research avenues for digitalised interventions~\citep{Stade2024}.

Yet, despite the excitement that surrounds them, there are still specific barriers to entry that stem from a lack of exchange and mutual understanding between the technical community that develops \acp{LLM}, and the mental health community that may benefit from them in clinical practice.
On the one hand, there is a substantial gap in the technical understanding that psychologists and psychiatrists might have of these modern-day systems, as well as a reluctant attitude or fear of misuse~\citep{VanDaele2022online}.
On the other hand, an equally dangerous gap extends to the technology side, as professionals might overlook the science of psychology, the needs of clinicians, and the potential risks of their tools~\citep{monteith2022expectations}.
%that's a cool paper on downsides of AI in mental health, which can go along with the automation bias and deskilling due to AI usage
In the present contribution, we attempt to bridge one side of this divide, namely that of a potentially limited understanding on the side of psychotherapy researchers and practitioners regarding the capabilities -- and potential shortcomings -- of those models. 
We also aim at improving the confidence of this audience in using these novel technologies in their clinical practice, and better understanding how unsolicited use by the patients themselves (through their own initiative) might impact their self-help strategies.
% This goal is already reflected in our section headings, where we attempt to bridge technical jargon with an easily-understandable explanation.
%JL:aiming at an informed perspective, potentially a souvereign usage of novel technologies in clinical work and better understanding of what and how their patients may already use as self-help strategies.
%JL: just a comment: we also know, that many patients use such services already, e.g. have conversations with siri or chatgpt, but would not tell their therapist about it, since it feels like a "betrayal" for them... In addition, many therapists are somewhat "scared" or uninformed and would not ask their patients about their usage of e-health service or tehcnology... I tried to capture this above, but I'm not happy yet with the formulation.
To that end, we present a short, concise overview of their basic operating principles with as little technical jargon as possible.
Crucially, we underline some of the fundamental assumptions behind the workings of those models, with an emphasis on those that are relevant for the domain of mental health.

In doing so, we primarily focus on the use of \acp{LLM} (and chatbots in general) for performing language-driven interventions \emph{(semi-)autonomously}.
While we briefly summarise potential application areas for \acp{LLM} in \cref{sec:appl}, it is worth to keep in mind the overall direction of the field -- major research efforts are currently directed towards improving the use of \acp{LLM} as general-purpose tools that can jointly handle conversation and text analysis.
The goal of this research agenda is encapsulated in the use of the term ``foundation models'' to describe \acp{LLM}, i.\,e., as models that serve as the foundational building blocks for a variety of applications~\citep{Bommasani21-OTO}.
%YT: I am not a fan of this decision and not so sure whether this is a good idea. Whey don't we provide a general overview how LLM Psychotherapy applications could be described (e.g., based on https://karger.com/ver/article/32/Suppl.%201/64/835090/The-Next-Generation-Chatbots-in-Clinical). Based on this general overview. I think highlighting the realm of tasks would also be a good idea.
%AT: how about we label it "language-driven" interventions?
%JL: but if you use images as well? (e.g. create an image based on your feelings?)
%AT: from the patient? this would mean turning this into a "multimodal" foundation paper, which Björn also suggested, but I am hesitant since we don't have enough space to do that

% This excludes other potential uses of \acp{LLM} in the broader psychology practice, such as summarising notes or running a more interactive form of \ac{EMA}.
% We also do not differentiate between fully autonomous agents that can carry an entire psychotherapy treatment by themselves (as promised by technology companies) and agents that `merely' support standard practice by carrying out semi-independent sessions under the broader guidance of a licensed psychotherapist.

In all cases, we focus on the role of \acp{LLM} as a human-empowering -- rather than human-substituting -- technology with the ultimate goal to \emph{improve patient outcomes} and restrict ourselves to an out-patient setting.
Our discussion remains agnostic to how that improvement is defined or measured (e.\,g., by means of a standardised assessment), how the treatment is prescribed (e.\,g., by a consulting psychotherapist), or which school of thought is followed (e.\,g., \ac{CBT}, psychodynamic approaches, or other).
Instead, we focus explicitly on \emph{how} the tasks related to this goal will be executed.
Moreover, we discuss on their potential limitations and how those may be addressed before they are deployed in a clinical setting.
Within this scope, we next introduce potential applications for \acp{LLM} in the next section, before discussing their inner workings.

\section{Applications for \acp{LLM} in mental health}
\label{sec:appl}
\begin{figure}[t]
    \centering
    \includegraphics[width=\textwidth]{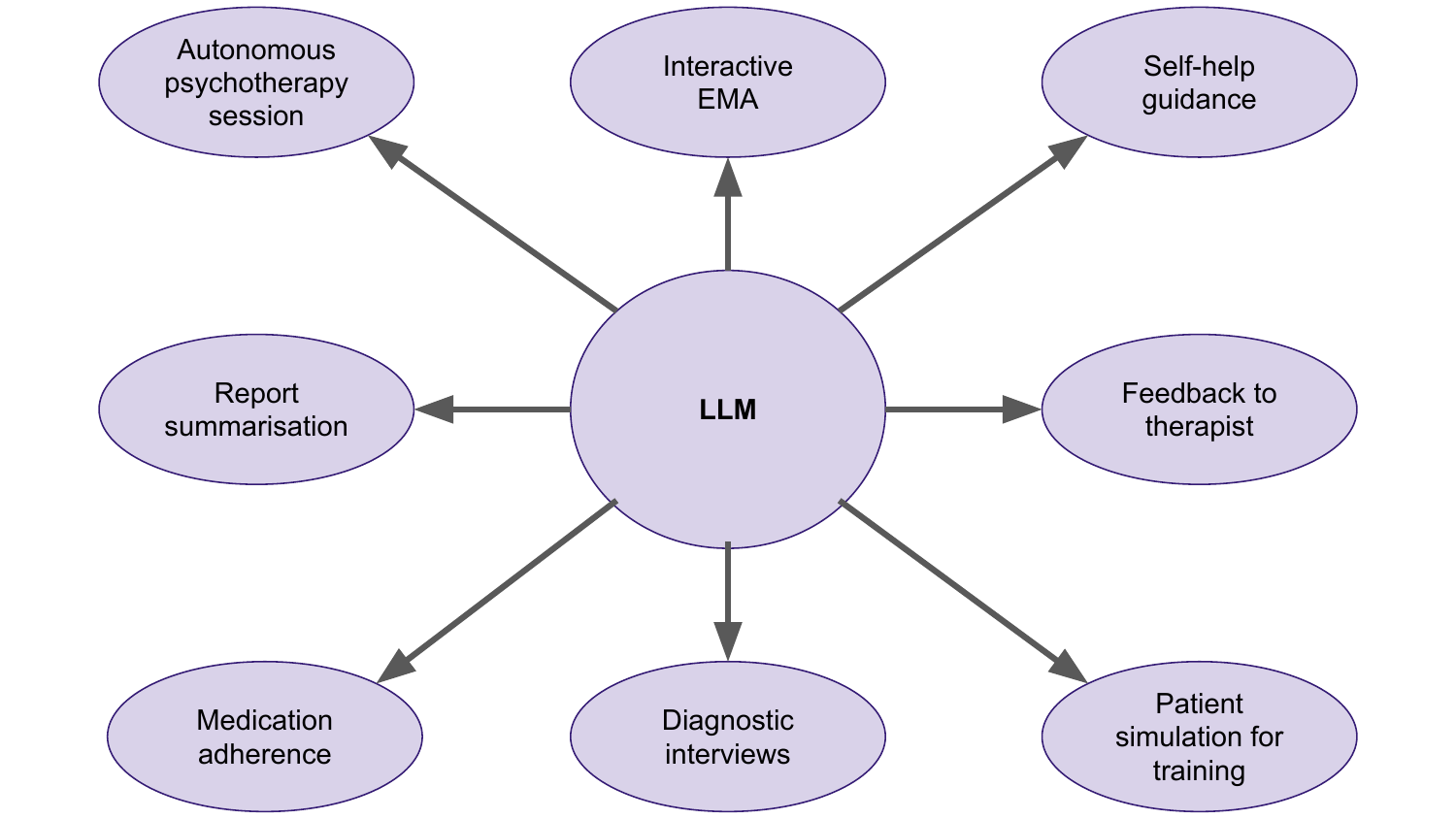}
    \caption{
    Overview of potential large language model (\ac{LLM}) applications in mental health, loosely inspired by the review of \citet{Bendig22-TNG}.
    Psychologists may use an \ac{LLM} to offload mundane tasks, such as note or report summarisation, or to obtain feedback regarding a particular therapy plan or even a single therapy session.
    Further, they might use \acp{LLM} to \emph{simulate} patient responses for the purposes of training.
    With respect to patients, \acp{LLM} can interact with them in different degrees of autonomy, ranging all the way from a fully-fledged, psychotherapy session, to mere reminders to the patient regarding medication.
    We note that some of those tasks can be achieved using more traditional software.
    However, \acp{LLM} provide both new avenues for improving efficiency, as well as offering a uniform platform that can achieve multiple tasks concurrently.
    }
    \label{fig:apps}
\end{figure}

\cref{fig:apps} shows a list of potential applications for \acp{LLM} in the daily practice of mental health researchers and practitioners.
We draw some inspiration from the recent review of \citet{Bendig22-TNG}, which covers existing applications of chatbots, and extended with tasks that can be performed by \acp{LLM}, but go beyond that to showcase a greater variety of possible applications. 
%yt: is this true? Bendig et al reviewed 6 pilot studies in the field. IMO the here outlined applicaitons as well as the general split regarding therapist-facing and patient-facing tasks is beyond the orginal paper. Also, the figures does not relate the 5 areas of characteristics (i.e., technical implementation; goal and endpoints; performance; areas of use; uderlying appraoch). Do we need to temper the wording here? IMO, this currently reads like "hey, we basically copied Bendig et al - please read the orginal paper", but this is not the case. 
We broadly distinguish between \emph{therapist-facing} and \emph{patient-facing} tasks.
Therapist-facing tasks are done ``in the background'', i.\,e., models do not interact with patients.
A therapist may use an \ac{LLM} to assist with note taking or summarising patient-therapist conversations~\citep{Kundan21-GSN}, or even use it as a \emph{role-play} substitute for actual patients in order to prepare themselves for an upcoming psychotherapy session, an application that is being considered by some authors~\citep{Qiu24-IAS}.
Moreover, they might use an \ac{LLM} to `score' their `performance' in a psychotherapy session~\citep{chaszczewicz-etal-2024-multi}; here, we do not mean using \acp{LLM} as an objective gauge of performance, but rather as a tool that can interactively provide feedback or suggest alternatives to the conversation that already took place.

Patient-facing applications instead feature a direct interaction of an \ac{LLM} with a patient.
This can be in the context of: 
a meditation practice;
adherence to a medication regiment;
self-management of symptoms via personalised messages~\citep{meyerhoff2024small};
cognitive restructuring \citep{10.1145/3613904.3642761};
mood-tracking \citep{schueller2021understanding};
journalling support \cite{10.1145/3613904.3642937};
or \ac{EMA} -- all in a more interactive, personalised fashion than is currently done using digital applications (i.\,e., by substituting a dry reminder with a contextualised suggestion)~\citep{seizer2024primer}.
Finally, we see \ac{LLM} agents that can autonomously run a psychotherapy session as the natural next step for current research, with the ultimate goal of developing entirely novel methods of psychotherapy and digital interventions in general. %yt: I would say the ultimate end-point are full in-silico trials of LLM-therapists treating LLM-patients and by doing so developing new psychotherapy methods, which worked based on thousands of in-silico sessions ;)
While their recent introduction has not allowed for large-scale studies, the impressive conversational capabilities of \acp{LLM} point to a future where they will be utilised as autonomous `psychotherapists', whether by initiative from the patients themselves, or after consultation with their therapist.
Nevertheless, the recent systematic review of \citet{Li2023} shows a significant reduction of depression symptoms (Hedge’s g 0.64 [95\% CI 0.17–1.12]) and distress (Hedge’s g 0.7 [95\% CI 0.18–1.22] when using conversational agents to perform an intervention, highlighting the promise of this research direction.
%JL: "not there yet" may not really be the case; see study of \citep{Li2023}, :" The meta-analysis revealed that AI-based Conversational Agents significantly reduce symptoms of depression (Hedge’s g 0.64 [95% CI 0.17–1.12]) and distress (Hedge’s g 0.7 [95% CI 0.18–1.22]). These effects were more pronounced in CAs that are multimodal, generative AI-based, integrated with mobile/instant messaging apps, and targeting clinical/subclinical and elderly populations." 
%maybe switch to .."Rising evidence of the positive effects on well-being of conversational agents \citep{Li2023} and the impressive conversational capabilities of \ac{LLM} point to a future where they will be utilised as autonomous `psychotherapists', whether by initiative from the patients themselves, or after consultation with their therapist.

\begin{figure}[t]
    \centering
    \includegraphics[width=\textwidth]{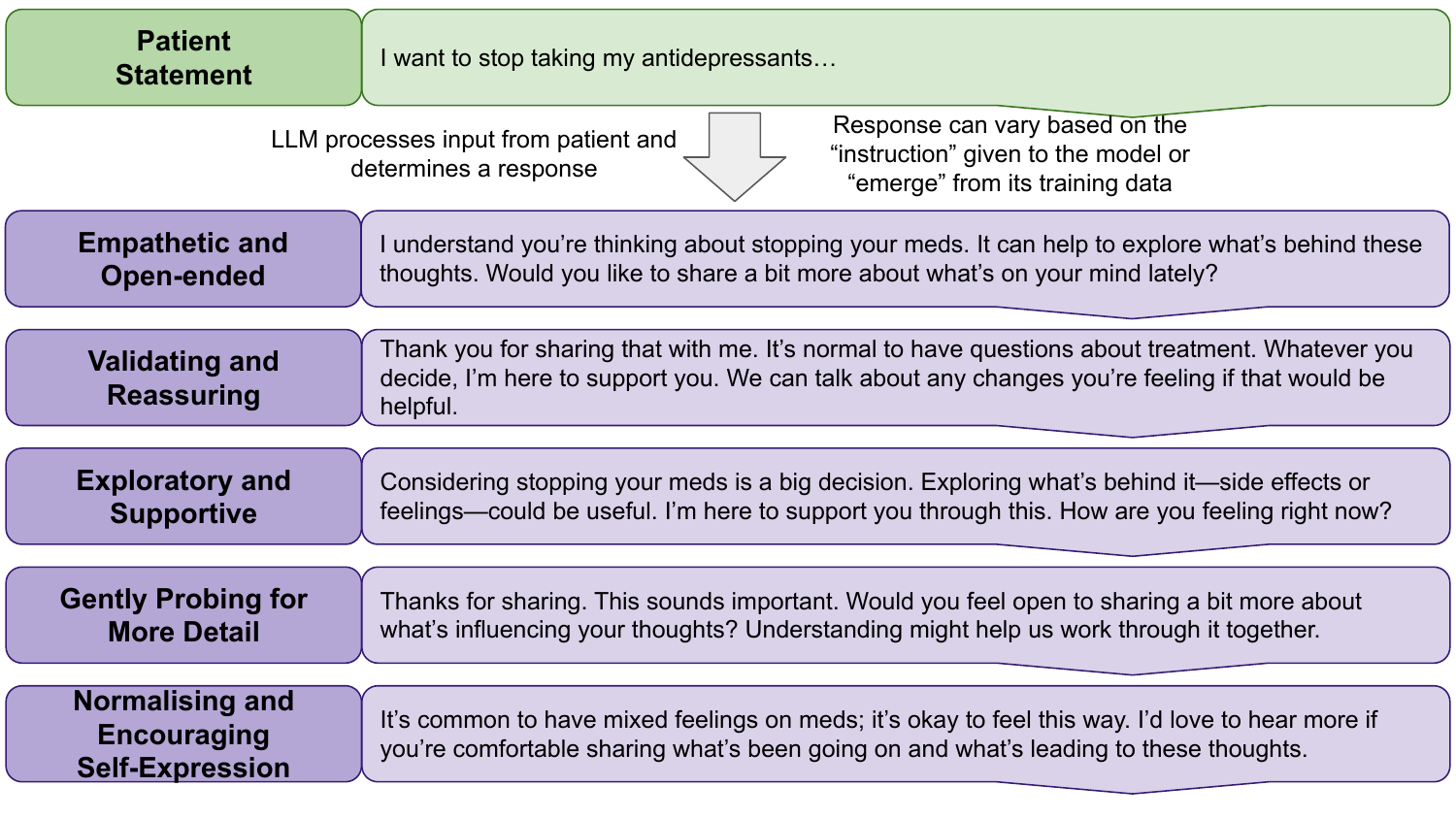}
    \caption{
    Example of a hypothetical \ac{LLM} response to a patient's intent to quit their antidepressant medication.
    In bolder font, we show alternative \emph{instructions} given to the model, which condition the type of its response.
    These instructions can be used to guide the \ac{LLM} towards particular types of responses that are conducive to a specific therapeutic plan.
    }
    \label{fig:example}
\end{figure}

An example of a hypothetical response given by an \ac{LLM} agent to the patient's intent to quit their antidepressant medication is shown in \cref{fig:example}. %yt very nice figure
The output can be conditioned to conform to different specifications that may be contingent on the patient's history and a specific therapeutic plan.

%BS: in the figure: unlabeled --> unlabelled
%BS: in the figure "Unsupervised pretraining" is "Self-supervised pretraining" - I understand that you want to avoid tech jargo, but unsupervised is another tech word and the wrong one - that's clustering, not self-supervised learning...
\begin{figure}[t]
    \centering
    \begin{tabular}{c|c}
        \includegraphics[width=.5\textwidth]{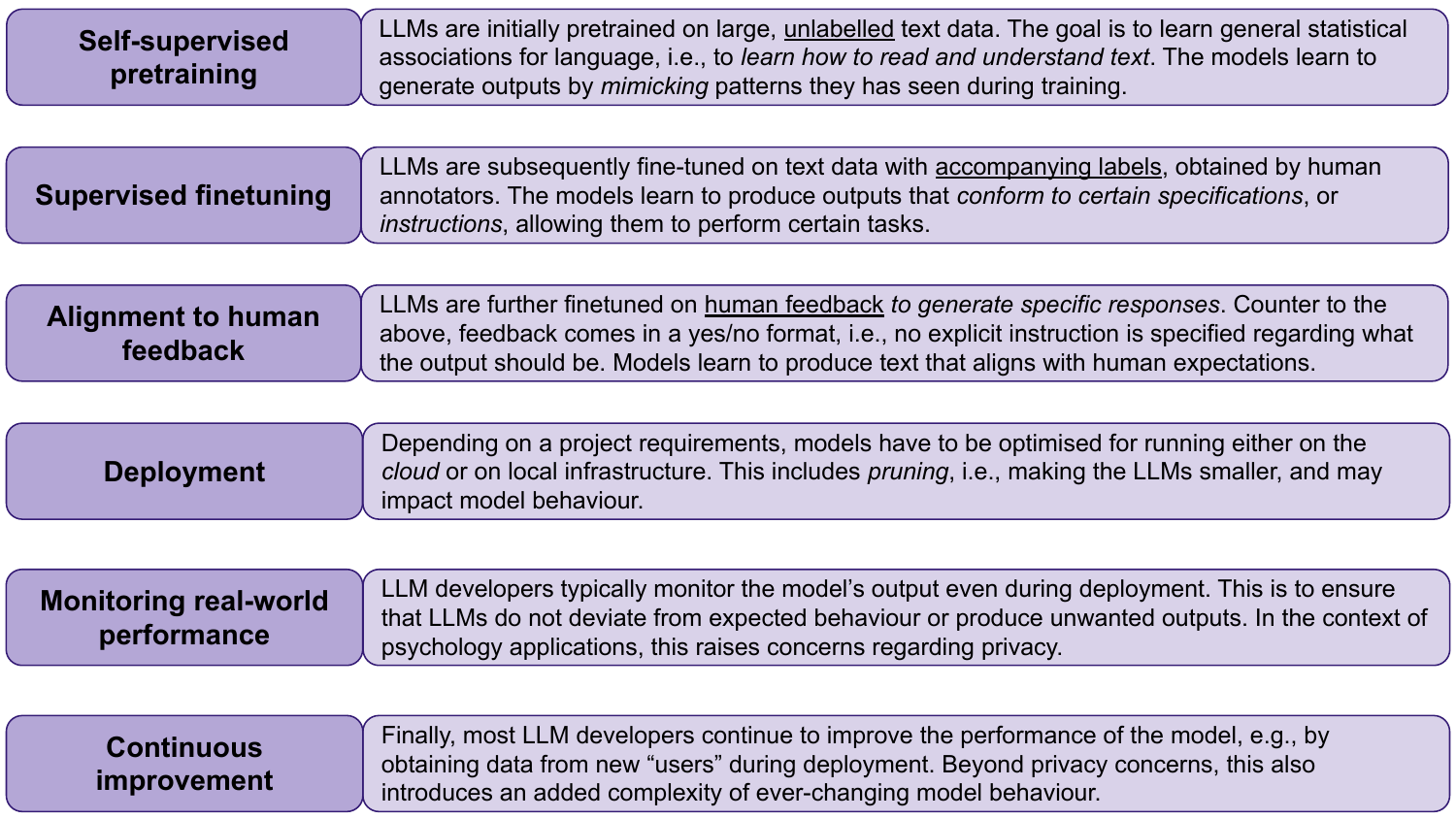} &  \includegraphics[width=.5\textwidth]{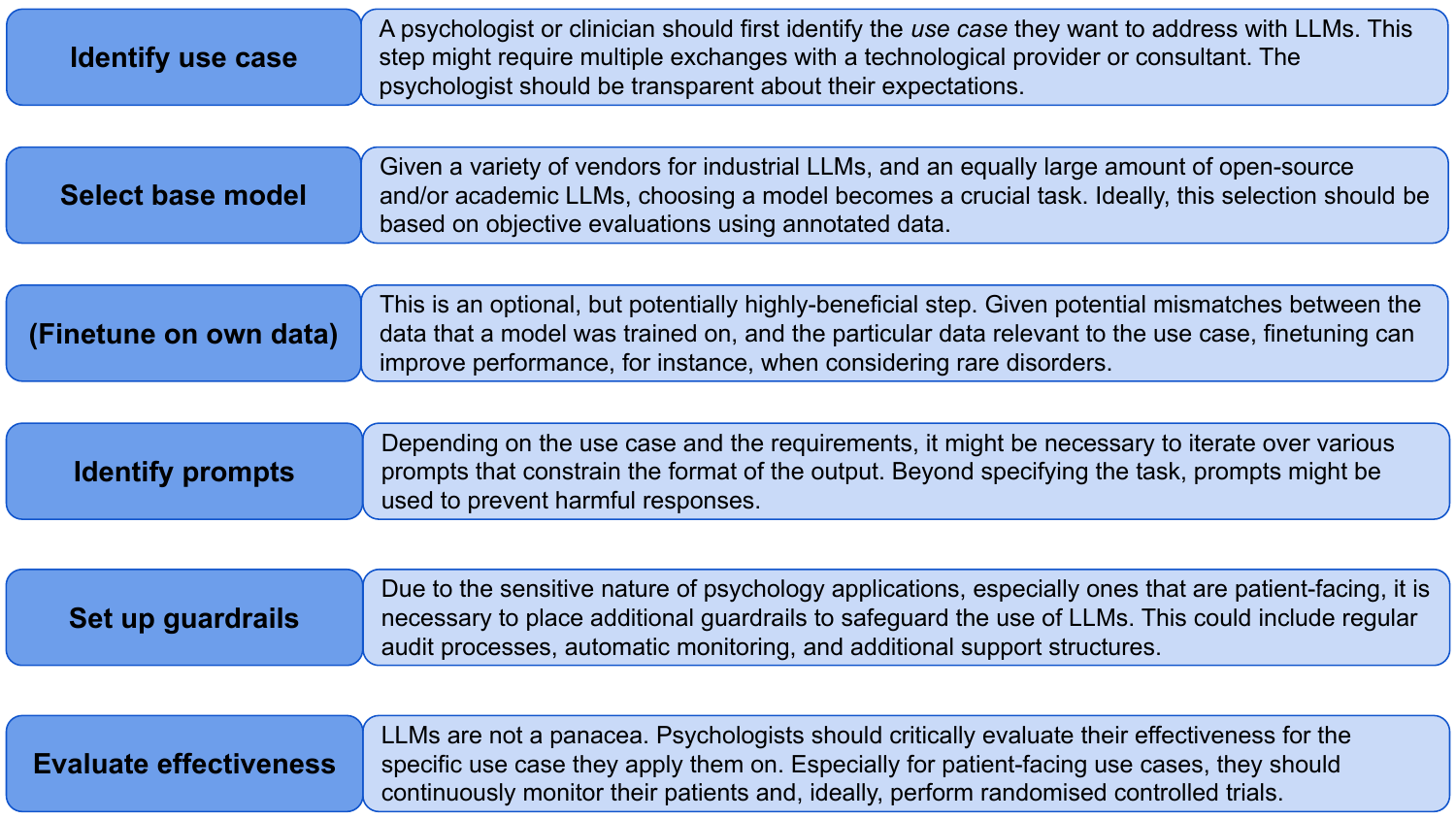}\\
    \end{tabular}
    \caption{
    Typical workflow for \ac{LLM} model developers (left) and proposed workflow for psychologists who wish to adopt \acp{LLM} in their practice (right).
    Developers follow a multi-stage training process (see \cref{sec:train}) to prepare their \acp{LLM}, and often continue to monitor and improve them even after deployment.
    Psychologists should carefully select the use case they want to automate using \acp{LLM}, conduct a preliminary evaluation of alternative models, and establish rigorous criteria for monitoring the effectiveness of \acp{LLM} with respect to patient outcomes.
    Moreover, the more technically versed can optionally finetune models on their own data, a process which can improve performance.
    }
    \label{fig:workflows}
    %JC - Is there scope to discuss (or mention a bit) where developers and psychologists should work together? For example, developers' responsibility in communicating the intended uses of their tool, limitations, what the evaluations mean; psychologists' expertise should inform how developers approach designing and implementing these models and evals; what happens when monitoring shows that models fail or exhibit bias (e.g. lets say LLM performance on younger patients are better than older onces), what happens then? 
\end{figure}

%BS: Please assure that all images can be zoomed in w/o turning into pixels for the fonts.
Given our intent to bridge the gap between psychologists and model developers, we additionally present the typical workflow of \ac{LLM} model developers in \cref{fig:workflows} (left), as well as our recommendations for how clinicians may go about integrating \acp{LLM} in their practice.
Naturally, we expect the two communities to heavily interact in the development of \acp{LLM} that translate to clinical practice by jointly identifying shortcomings of existing models and working together to overcome them.

\noindent
\textbf{What do developers do with \acp{LLM} --} Model developers go through a 6-step process of model creation and validation.
The first three steps correspond to how a model is trained and are further discussed in \cref{sec:train}.
The last three steps correspond to a continuous cycle of validation and improvement.
Note that, for commercial offerings, this includes \emph{using data obtained by users of the \ac{LLM}}.
This has obvious privacy implications for mental health patients and their therapists, and should be taken into consideration when using such models.
In the case of open-source models, which can be run on local infrastructure, this is not necessarily the case.

\noindent
\textbf{What should psychologists do with \acp{LLM} --} We also tentatively prescribe a 6-step process for psychologists who wish to adopt \acp{LLM} in their clinical practice.
While far from a comprehensive suite of guidelines, we consider it necessary to highlight \emph{how} one should go about using these models before discussing how they are made. %yt IMO, the six steps do not sufficiently become clear. What are the six steps? At least I can only see two explicitly stated steps. Consider a numbered list.
The first step should be to identify the context in which an \ac{LLM} should be applied (perhaps inspired by the potential applications we show in \cref{fig:apps}).
This will determine the requirements for model performance (e.\,g,. note summarisation is less critical than giving feedback to the patient).
Following that, a psychologist should identify a \emph{base model} that has been pretrained and is publicly-available.
Contemporary examples include, among others, ChatGPT (commercial), Claude (commercial), or 
%BS: I thought it is spelt LlaMa, but Llama seems to be used by meta as well - please check.
Llama (open-source; trained by a commercial entity).
Ideally, the selection should be based on a rigorous evaluation of model capabilities \emph{using data that is available to the prospective user}.
Note, again, the obvious privacy implications -- this evaluation should be done without unwittingly exposing the data to an external entity, i.\,e, the model should be used within the local computer infrastructure.
Optionally -- and assuming a certain amount of technical expertise -- the selected model can be finetuned on own data.
While not strictly necessary, this process can improve model performance by adapting to the local domain (for instance, to account for idiosyncrasies of the patient population that appears in a particular practice).
After selecting a model and finetuning it, it is necessary to identify the \emph{prompts} required by the model to achieve a particular task in a particular way.
While prompts will be explained in \cref{sec:prompt}, they are similar to the instructions shown in \cref{fig:example}.
Finally, any clinician should consistently and critically evaluate the performance of the model, and set up additional guardrails for its use in clinical practice. 
For instance, we do not recommend completely offloading diagnosis to a model.

\section{Basic operating principles of \acp{LLM}}
\begin{figure}[t]
    \centering
    \begin{tabular}{c|c}
        \includegraphics[width=.5\textwidth]{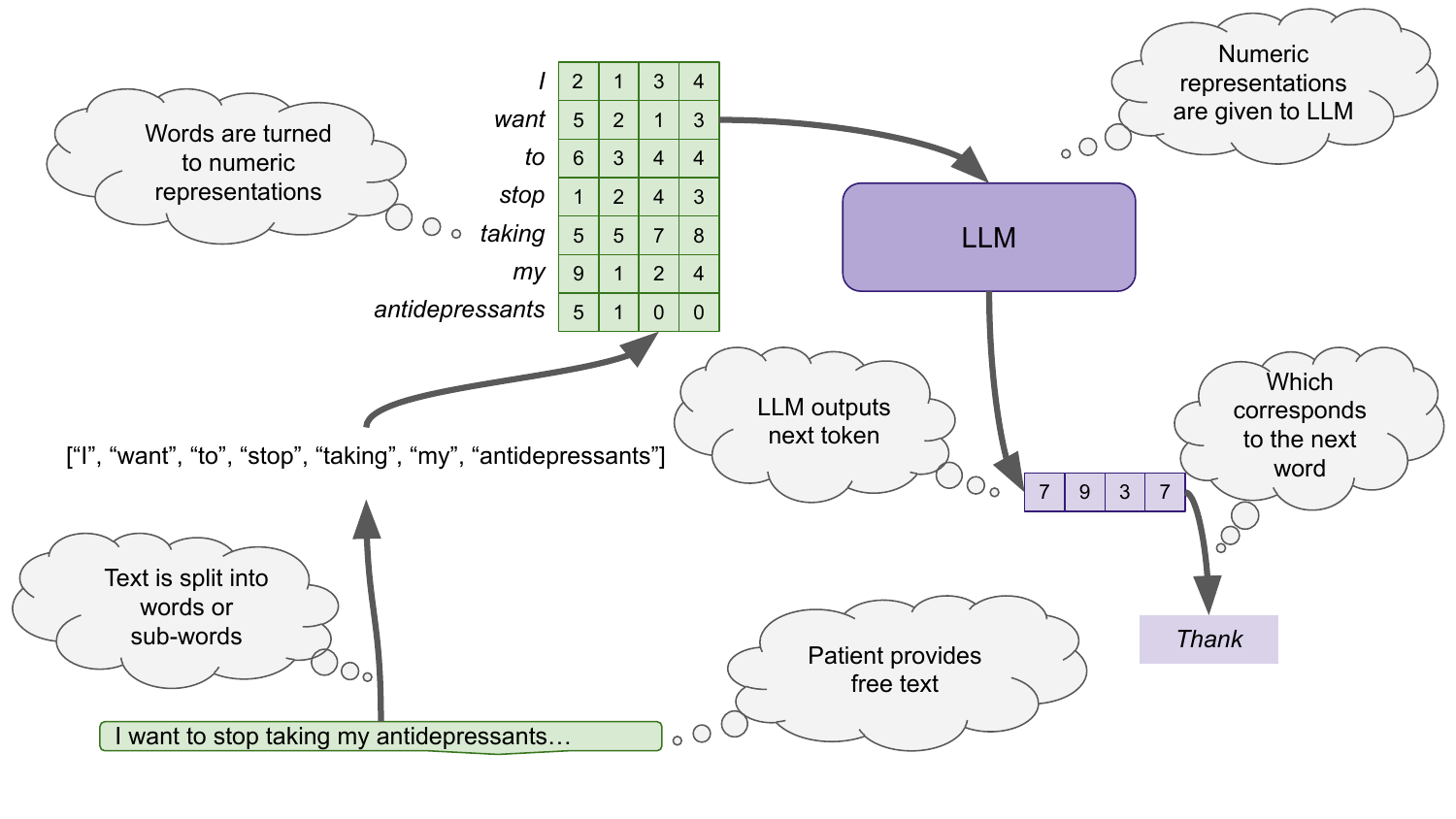} &  \includegraphics[width=.5\textwidth]{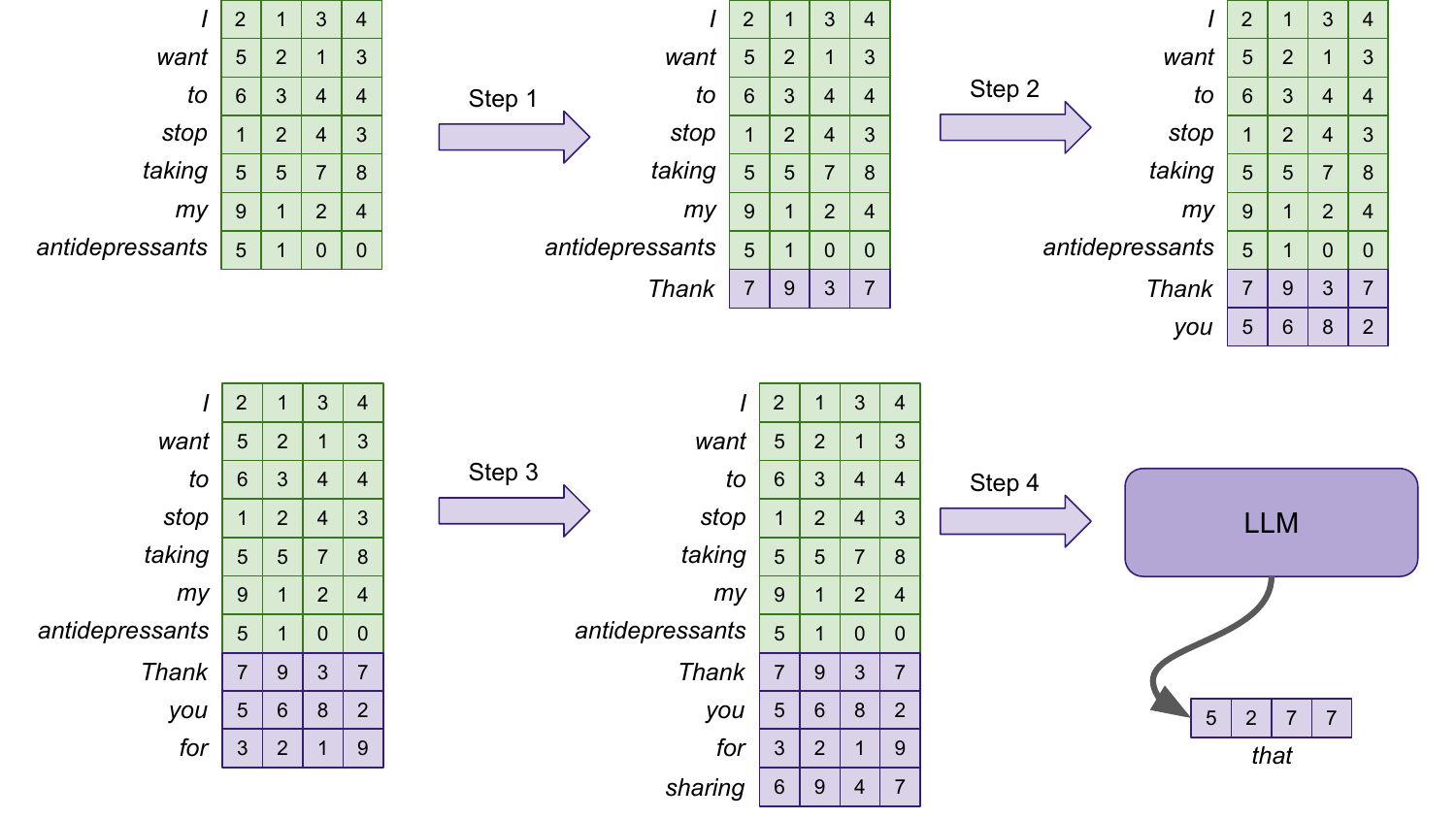}\\
    \end{tabular}
    \caption{
    Overview of \emph{autoregressive} output generation by an \ac{LLM}.
    The left-side panel shows a schematic of the major steps in an \ac{LLM} workflow: 
    a) the input text is split into words or sub-words (see \cref{sec:tokens}) and subsequently converted to a numeric representation;
    b) this numeric representation becomes the input to the \ac{LLM};
    c) the \ac{LLM} outputs the \emph{most likely} next word in the sequence (as a numeric representation that can be interpreted as a word).
    The right-side panel shows the initial step of this autoregressive process for generating one of the responses shown in \cref{fig:example}.
    Note that the output word at each step becomes part of the input for the next step (hence the name autoregressive).
    }
    \label{fig:llm-overview}
\end{figure}

On a high conceptual level, \acp{LLM}, like all other machine learning models, aim to \emph{predict a suitable output given an input}.
In the specific case of \acp{LLM}, they are processing a text input (so called \emph{prompt}) to generate a text output (so called \emph{completion}).
In the simplest possible terms, all this means is that they are able to generate coherent text;
as will be explained below, they do so in a so-called autoregressive fashion by producing one \emph{token} at a time -- \emph{token} is simply the technical term used to denote word or sub-word units.
% Each \ac{LLM} features its own \emph{tokenisation} step, which splits input text into discrete components.
% In theory, this could be the entire set of words found in a given language.
% However, given the relative low frequency of some words, and the ever-changing nature of language, \acp{LLM} typically feature a more limited vocabulary, and split out-of-vocabulary words into smaller, sub-word units.

Their output depends on the preceding inputs; usually, these inputs are also texts; though multimodal models are increasingly being developed, we will focus here on text-to-text conversations.
In the case of \acp{LLM} used for conversations, these preceding inputs comprise both the user input and previous outputs by an \ac{LLM}; in other words, every new output depends on the insofar conversation (within limits; please also see below).

%YT: I think a example figure would be great: prompt "I am feeling depressed" --> LLM Psychotherapist --> Completion examples: 
% 1. Empathic and Open-Ended: "I'm really sorry you're feeling this way. Can you tell me more about what your depression feels like and how it's affecting your life? We can work together to understand what’s going on and find ways to manage it."; 
% 2. Validating and Reassuring: "Thank you for sharing that with me. Feeling depressed can be overwhelming, but you're not alone in this. It's important that you're here, and together we can explore what might be contributing to these feelings and how we can improve things.";  
% 3. Exploratory and Supportive: "I'm really sorry to hear you're struggling. When did you start feeling this way? Are there any specific events or thoughts that seem to trigger your feelings of depression?"; 
% 4. Gently Probing for More Detail: "That sounds really difficult. What do you mean by depressed—are you feeling sad, empty, or lacking energy? Understanding exactly how you're feeling can help us figure out the best way forward."; 
% 5. Normalizing and Encouraging Self-Expression: "It’s understandable to feel overwhelmed when you’re depressed. Many people experience this at some point, but it’s different for everyone. Can you describe more about what you're feeling emotionally and physically?"
% AT: see fig:example %YT: very nicely done.

Crucially, this process of language generation is statistical in nature.
To generate each output, an \ac{LLM} computes the statistical probability for each possible token in its \emph{vocabulary}; then, it outputs a highly-probable next token.
This process can be repeated \emph{ad infinitum} to generate sentences of arbitrary lengths.
The function which estimates the probability of the next token relies on a \ac{DNN}.
%JL: a example would be great here already? e.g. there is a high statistical probabilty of the token "depression" in case someone enters "sad, hopelessness, loss of motivation and energy"? Does this make sense?
An intuitive explanation of the overall process is shown in \cref{fig:llm-overview}.
In the next sections, we discuss how those \acp{DNN} are implemented (\cref{sec:arch}), how they are trained (\cref{sec:train}), and how they can be controlled (\cref{sec:control}).

\subsection{Neural architecture}
\label{sec:arch}
As mentioned, \acp{LLM} are essentially \acp{DNN} trained to approximate a probability function -- and, in particular, the probability of one particular token appearing next in the sequence of tokens that the \ac{LLM} is currently generating.
We therefore begin by discussing the process of tokenisation in \cref{sec:tokens} and the paradigm of \emph{autoregressive modelling} in \cref{sec:autoreg}.
Following that, we present the \ac{DNN} architecture that underpins most modern \acp{LLM} in \cref{sec:trans}, and, finally, the computational limits of \acp{LLM} in \cref{sec:context}.

%YT: I think that somewhere a high-level overview of the whole process would  be goood: Whart are the conceptual steps a input runs through until an output is generated: When the text input "I am depressed" is fed into an LLM, it first undergoes tokenization, breaking the sentence into smaller units, often as ["I", "am", "depressed"]. Each token is then mapped to its corresponding word embedding, a high-dimensional vector representing the token's semantic meaning. These embeddings capture the context and relationships between tokens. The model processes these embeddings through multiple layers, applying attention mechanisms to understand context, then generates an appropriate output based on the task, such as predicting the next word, answering questions, or completing the sentence.

\subsubsection{Tokenisation and embedding -- Turning text into numbers}
\label{sec:tokens}
Sentences are made of words laid one after another, each word comprising some combination of the letters in the language's alphabet (plus punctuation, white spaces, numbers, or special characters). %JC - to be overly pedantic, there are also whitespaces, numbers, special characters, and characters from other languages/scripts. 
Digital systems represent each of those characters with a unique numeric representation (e.\,g., in ASCII or Unicode).
In the simplest case, these numeric representations can form the input to an \ac{LLM}.
However, character-level representations are highly inefficient, as one must first learn the associations between different characters to understand individual words.
For this reason, most \acp{LLM} use a higher semantic unit as their input -- which is referred to as \emph{tokens}.

On the other extreme to characters are words.
Individual words offer an alternative as input to \acp{LLM} -- all one needs to do is represent each individual word with a unique number.
Why then \emph{tokens} and not \emph{words}?
The Oxford English Dictionary boasts a collection of ``over 500,000 words (and phrases)'', and this excludes the different forms that verbs and nouns can take (e.\,g., singular vs plural, verb conjugation, tense).
As \acp{LLM} are required to be `open-world' chatbots, i.\,e., to converse about any potential topic, they should be able to process, and potentially generate, every single one of those words; accordingly, every word should be given its own numeric representation.
However, the size of the desired vocabulary is prohibitive for the computational capabilities of contemporary systems, especially since they are trained to support multiple languages~\citep{Jean15-OUV}.
For instance, even the Llama-3 models have a vocabulary of `only' 120,000 tokens~\citep{Dubey24-TL3}, which does not suffice to cover the entire English language. %yt: in this paragraph you already use tokens as terminology before the formal definition and explanation what a token is.

A middle ground is found by splitting words into sub-word units -- the so-called \emph{tokens}~\citep{Sennrich16-NMT}.
In practice, most `frontier' \acp{LLM} have sufficiently large vocabulary spaces that cover most common words.
However, even the state-of-the-art GPT-4o and Llama-3 models~\citep{Dubey24-TL3} -- two of the most advanced models available to the public -- will split some common words to subwords\footnote{One can experiment with their tokeniser here: \url{https://belladoreai.github.io/llama3-tokenizer-js/example-demo/build/} and \url{https://platform.openai.com/tokenizer}.}; Llama-3, for instance, has a vocabulary size of 128,000 tokens~\citep{Dubey24-TL3}.
For instance, the word ``antidepressants'' is split into ``antidepress'' and ``ants''.
This influences how these particular \ac{LLM} will deal with sentences like ``I want to stop taking my antidepressants''; we return to this later.
%JL:that's really brilliantly explained!
% The reason is that the last 
% FFNN: 2617245696 (per layer) --> 329,772,957,696
% ATT: 553648128 (per layer) --> 69,759,664,128
% 2,097,152,000 OUT 
% 8,192,000,000
% \citet{Sennrich16-NMT}

% Accordingly, there are many different ways to perform tokenisation (e.\,g., text: ``I want to stop taking my antidepressants''; Byte-Pair Encoding: [``I'', `` am'', `` feeling'', `` depressed'']; WordPiece Tokenisation: [``I'', ``am'', ``feeling'', ``de'', ``\#\#pressed'']; SentencePiece Tokenisation: [``\_I'', ``\_am'', ``\_feeling'', ``\_de'', ``pressed'']. %YT: I stopped expanding this, since I am not sure what will be in appendix A. However, the concept of tokens or even vector embeddings does not become clear for psychologists here. 
% \ak{I think that for a naive reader, it is not fully clear why tokenization as "breaking words and sentences into small units" is related to "turning words into numbers" or "bring them into digitilized form". To bridge this gap we should introduce the notion of indexing a vocabulary --- we chop the text into basic building blocks called tokens; These tokens are then organized into a vocabulary; this way the text can be translated into a sequence of numbers - the tokens indices by the vocaulary.  }

% \textbf{Key considerations:} Tokenisation influences how words are processed and perceived by an \ac{LLM}.
% As we will discuss below, it is tokens that are generated one at a time, not words.
Following tokenisation, tokens need to be transformed into numbers.
In the simplest possible case, they are sorted alphanumerically, and an integer is used to represent each one -- this is referred to as the \emph{id} of that token.
However, one immutable integer representation per token is not enough to capture semantics -- this is why these initial encodings are transformed into high-dimensional \emph{vector} representations, known as token \emph{embeddings}.
These vectors are typically decimal numbers (note that our figures show integers for visualisation purposes) and, more importantly, are \emph{learnt from data}.
Typical examples include \emph{GloVe} embeddings~\citep{Pennington14-GGV} or \emph{word2vec}~\citep{Mikolov13-DRO}.
It is these vector representations that are given as inputs to an \ac{LLM} and processed by its inner layers to produce a prediction for the next output token. %yt: you loose the psychologists here. I would stronlgy recommend to include a visualization here or at least provide reference for a psychologist-understandable intoruction (e.g., https://www.cs.cmu.edu/~dst/WordEmbeddingDemo/tutorial.html ). 

\subsubsection{Autoregressive modelling -- Output generation}
\label{sec:autoreg}
The goal of an \ac{LLM} is to produce an output text that is \emph{appropriate} with respect to the past history of the conversations.
This is done by generating one token at a time until a limit is reached; this limit is decided either by the \ac{LLM} itself, as it has the option to output a so-called `end-of-sequence' token that signifies its turn is finished, or via some hardcoded rule (e.\,g., output at most 50 tokens).
Every new token that is generated is \emph{fed back} as input so that the subsequent token can be generated, and so on.
The process is kickstarted either by the past history of the conversation, a suitable prompt (\cref{sec:prompt}), or a combination of the two.
This workflow is illustrated in \cref{fig:llm-overview}.

The process we just described is often referred to as the \emph{greedy} variant of autoregressive modelling; it is considered greedy because at each step, it picks the most probable next token given all previous steps; however, this might not be the most \emph{appropriate} choice when the entire output sequence is taken as a whole.
Alternative forms of decoding are based on variants of \emph{beam search}~\citep{Vijayakumar18-DBS}, i.\,e., algorithms which explore different encoding sequences in parallel, and only select the most likely sequence after a few steps.
It is important to note that there is no explicit thought process that first `decides' what the appropriate response is, and then generates it; instead, \acp{LLM} follow a \emph{think-as-you-go} paradigm that formulates the response on a surface, lexical level without determining an explicit, communicative intent in advance.
By this, we do not mean to rule out any `higher-level' thinking done by \acp{LLM}, but rather to emphasise that this higher-level thinking can only arise as a byproduct of the training data (or be unlocked via prompting, see \cref{sec:prompt}), although the discussion on this is still ongoing~\citep{Bender20-CTN}.

% \ak{I think this paragraph is confusing, and even misleading. In essence it conflates two discussion - one, the discussion of the decoding or output generation mechanisms (and the fact thate this "think-as-you-go" paradigm diverge from what we imagine as the natural way of linguistic expression, i.e. having a full grasp of the 'communicative intent' or semantics before starting to compose the linguistic output word by word); and two, the discussion of generalization and dependence on training corpus. I think both are important discussions but they are distinct. The latter discussion concerns ML in general, and perhaps more suitable in Section 3.2 where we discuss training. Here we can keep pondering about the existence of "higher-level" thinking, and cite Bender & Koller, "Climbing towards NLU: On Meaning, Form, and Understanding in the Age of Data", which state a strong claim opposing the attribution of "understanding" to language models.  }
%AT: you're right, I split the two

\subsubsection{Transformers -- \ac{LLM} building blocks}

\begin{figure}[t]
    \centering
    \begin{tabular}{c|c|c}
        \includegraphics[width=.33\textwidth]{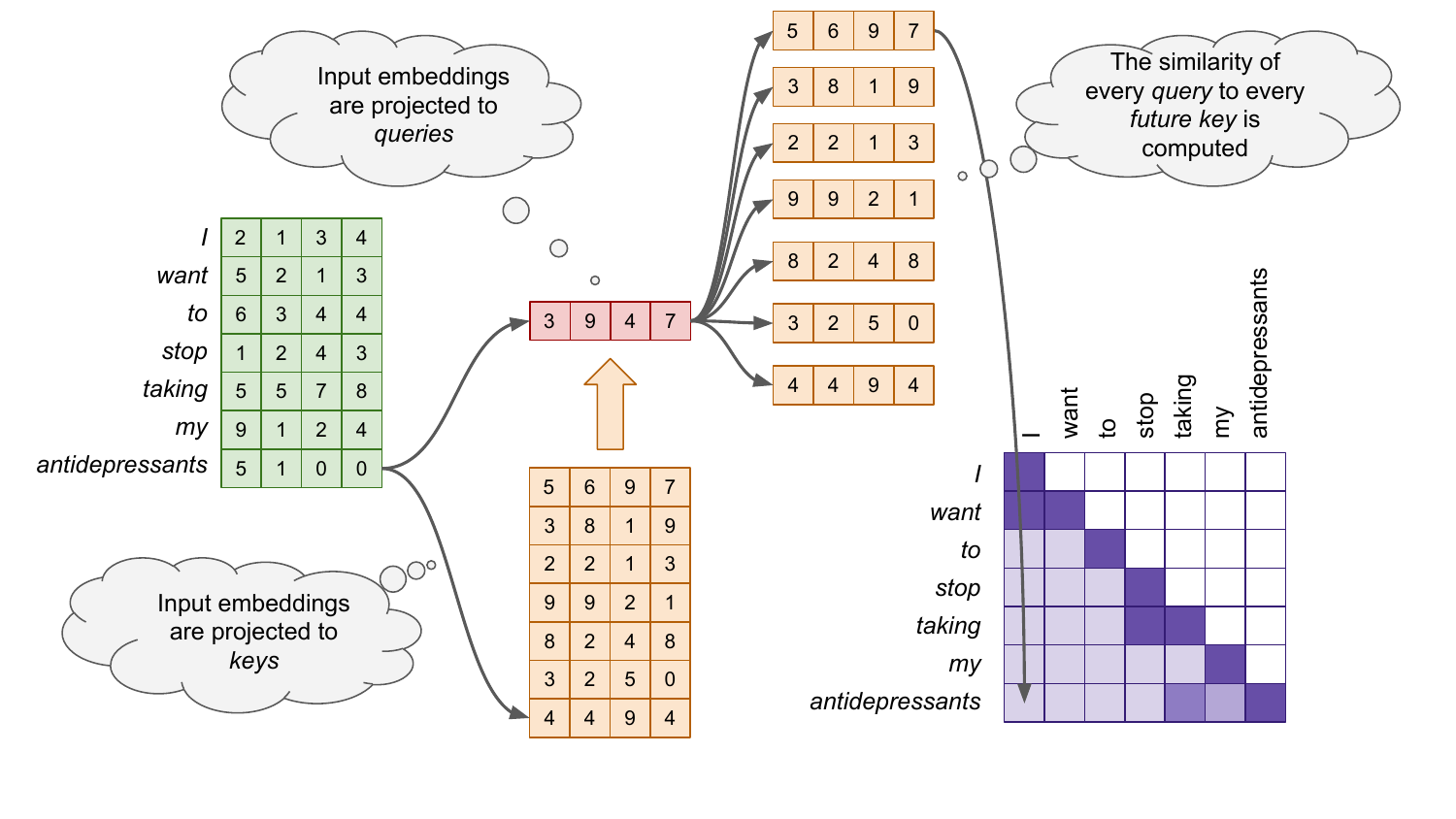} &  \includegraphics[width=.33\textwidth]{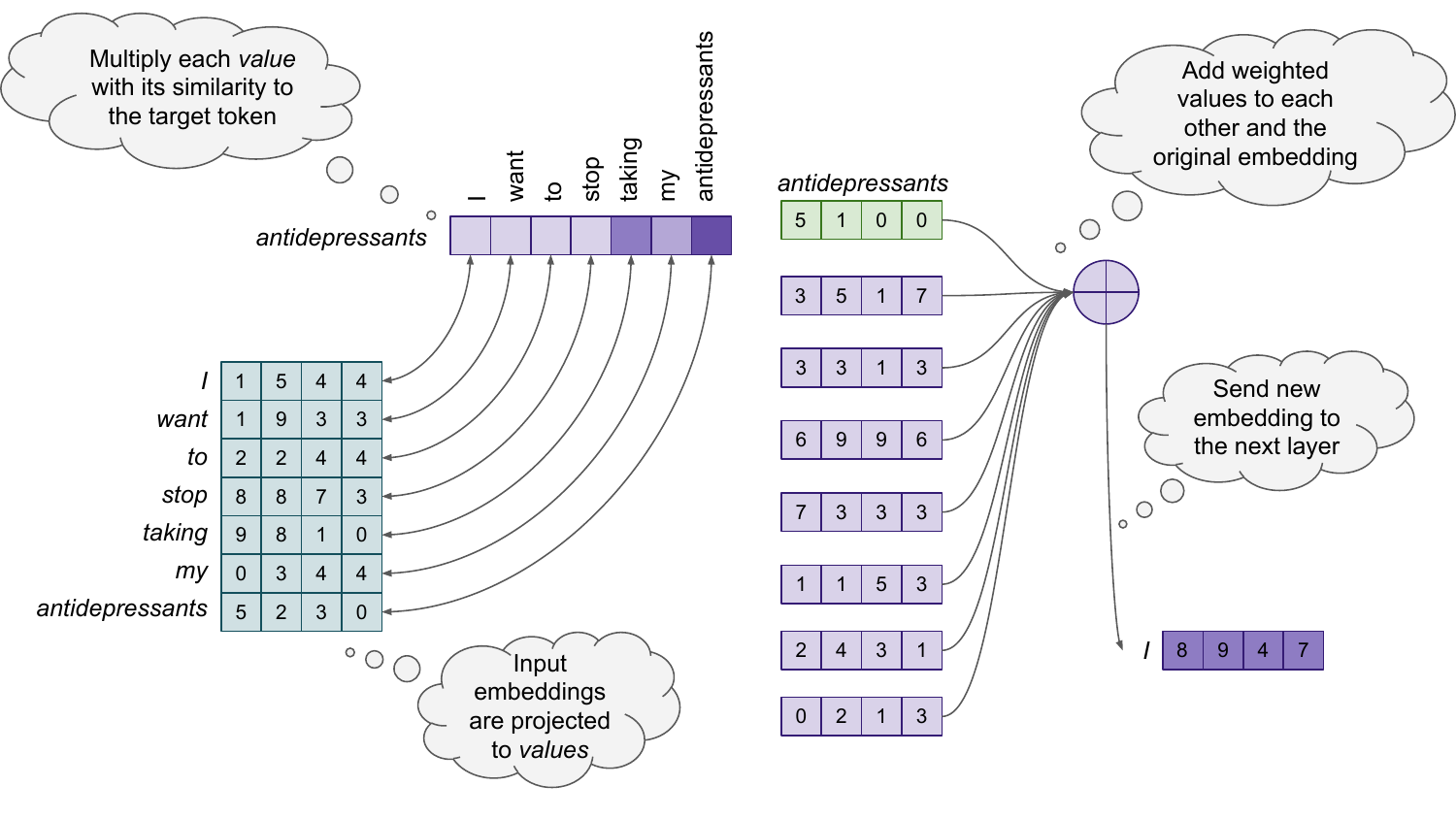} &  \includegraphics[width=.33\textwidth]{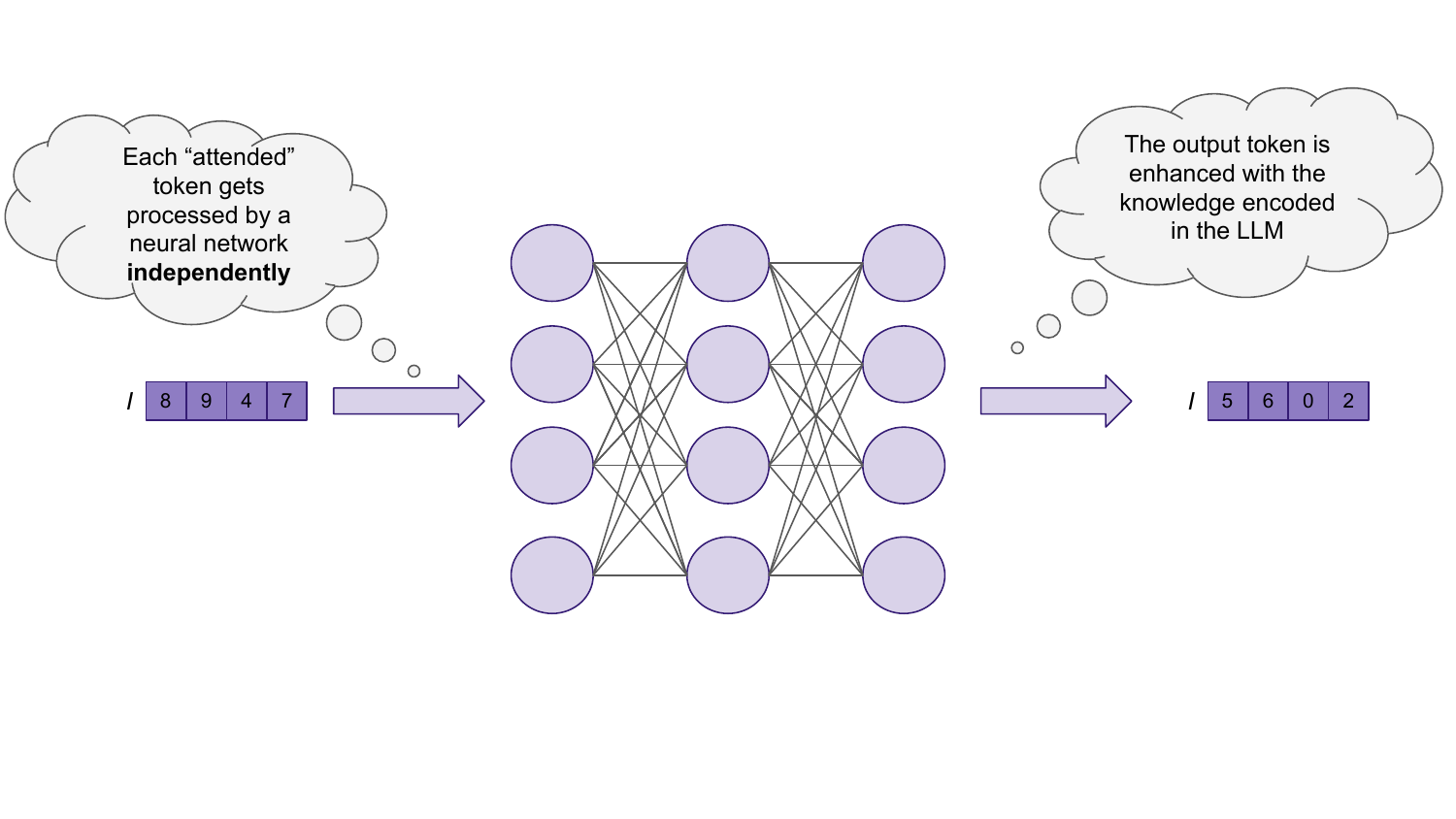}\\
    \end{tabular}
    \caption{
    Overview of the inner workings of the \emph{transformer} architecture which underpins most contemporary \acp{LLM}.
    The leftmost panel shows the inner workings of the \emph{attention} layer, a key component of the transformer architecture.
    The initial numerical representations of words (green table) are first linearly transformed into a specific \emph{contextualised representation} (orange table), which is then used to compute similarities across all words in the input (resulting in the purple matrix).
    The linear transformation is learnt from data and is used to specify different \emph{notions of similarity} (e.\,g., grammatic vs affective similarity; see text for more details).
    This is the first stage of the attention layer.
    The middle panel shows the second stage of the attention layer, which computes a weighted representation from each word, derived from its (contextualised) similarity with all other words in the input.
    In a nutshell, this \emph{transfers information} across all words in a sentence.
    In our example, this could be used to identify that the verb \emph{taking} corresponds to \emph{my antidepressants}; the representation of \emph{taking} would then be enhanced by the representation of \emph{my antidepressants}, thus placing this word in the correct context for this sentence; given that several of those layers are applied one after another, a later layer would then connect \emph{stop} with \emph{taking my antidepressants}, to finally identify the intent of the patient.
    Finally, the rightmost panel shows the workings of the \emph{multilayered perceptron} which processes the output of each attention layer; this further contextualises the words in the sentence with \emph{information learnt during training}.
    In our example, this could be used to identify that the phrase \emph{stop taking my antidepressants} is associated with negative connotations.
    }
    \label{fig:transformer}
\end{figure}

\label{sec:trans}
Most contemporary \acp{LLM} depend on the so-called \emph{transformer} architecture.
This is an architecture first proposed in \citet{Vaswani17-AIA}.
It is made of \emph{repeating blocks}, with each block comprising a \emph{self-attention} layer followed by a two-layered \emph{multilayered perceptron} -- with some additional normalisation layers in-between.
A high-level overview is shown in \cref{fig:transformer}.
We discuss the role of each component below. %YT: I think that the figures are placed not ideally throughout: interupt reading flow, appearing before the first reference to them etc.

\textbf{Self-attention:} Self-attention layers operate in the following way.
First, they compute a notion similarity between all tokens in the input; this includes the tokens that the \ac{LLM} has generated in its present turn, but also the entirety of the conversation up to this point.
This \emph{notion} of similarity is different for each self-attention layer and is determined by weights learnt during training; each layer can thus specialise to a different notion of similarity.
For example, one particular layer might specialise in semantic similarity, 
another for part-of-speech (i.\,e., an adjective that describes a particular noun will have the highest similarity with that noun compared to all other input tokens), and yet another for affective similarity.
Specifically, the embeddings of each token are projected twice using two projection matrices that are learnt during training; the outputs of those two are referred to as \emph{queries} and \emph{keys}. %yt: Psychologist will not know what projection is.
Following that, the similarity is computed between each query and key -- though only for keys that come after a query in the sentence, in order to ensure causality.
These similarity scores are known as the \emph{attention matrix}.

Finally, the embeddings of each token are further projected (using a learnable projection matrix) to a set of \emph{values}; these values are weighted with the corresponding attention weights and added to the original representation.
In this way, information is propagated from each one token to all other tokens \emph{weighted} by the respective similarity of each pair; this ensures that at the output of a self-attention layer, each token will carry the additional information that it needs from all other (previous) tokens.
An illustrative example is shown in the first two panels of \cref{fig:transformer}.
There, the similarity of the last word, \emph{antidepressants}, is computed with all other tokens in our example (left panel), and this similarity is then used as weight when adding the representations of those other words to the representation. 
%AT: I will put these in a figure:)
% (i.\,e., looking for synonyms, such as ``medication'' or ``pills'' for ``antidepressants'' and ``dropping'' for ``quitting'') %JL: e.g. job --> work/occupation, quitting --> drop/leave/..?
%JL: I know it adds a lot of words, but I think concrete examples help understand this process...
% In our example above (``I am thinking to stop taking my antidepressants''), a self-attention layer specialised in finding grammatical similarity might presumably connect ``quitting'' and ``job''; then the information from the two will be combined at the output of the layer; so the tokens previously representing just ``quitting'' and ``job'' will at the output of the layer represent a novel concept that interpolates between the two.
% This can be extrapolated to account for the rest of the tokens in the input (for example, by adding ``I'' so that it becomes evident that it is the patient who is thinking of quitting his*her job).
%yt: Honestly, I think this is still overwhelming for psychologist. I would recommend to break the figure down into separate ones and explain the steps with the help of these separate in figures step-by-step and in more detail. 

This is the mechanism that the model can use to `repair' the `damage' done by splitting ``antidepressants'' to ``antidepress'' and ``ants'' during tokenisation (see above).
By leveraging the context present in the rest of the sentence (as well as, potentially, previous sentences), the model will connect ``ants'' with ``antidepress'' and interpret them as the constituents of a singular word that has been inadvertently split, rather than interpret ``ants'' as actual ants and ``antidepress'' as a quantifier.

\textbf{Multilayered perceptron:} After self-attention -- which has enhanced the content of each token by adding information from each other token, a two-layered linear layer processes each token \emph{independently}.
These layers no longer specialise in carrying information across tokens, but rather on enhancing the information within a single token, and can be intuitively thought of as the more standard ``feature extraction'' process of classical \ac{NLP}, like the \ac{LIWC} features~\citep{Boyd22-TDA}.
When a token is passed through this type of layer, it is `endowed' with the additional world knowledge that the \ac{LLM} has acquired during its training.
For instance, the word ``stop'' in our example will be interpreted as a verb, rather than a noun, and, after it is combined with the rest of the context preceding it, will be interpreted to have a negative connotation.
% \todo[inline]{This needs to be fleshed out a bit more}
%JL: if this is for Psychologitsts, thex may not be familiar with feature extraction and NLP, just those that work with it maybe... %yt: fully agree.

\textbf{Key considerations:} We noted how the transformer architecture can be thought of in terms of two main components: one that cares for cross-token relationships and another that does within-token processing. %yt: this words should be appearing before and not just in the key considerations. IMO the key consideration should be a comprehensive summary of the key points, but not introduce new jargon and terminology. Maybe this section should be placed before the subheadings as a primer providing some meta-structure/guidance?
This is meant to emphasise -- again -- how all \emph{knowledge} exhibited by an \ac{LLM} is present in its weights.
Self-attention layers trigger this knowledge by connecting the different tokens to one another in a sentence whereas perceptrons then further contextualise this combined knowledge~\citep{Elhage21-AMF}.
Importantly, this functionality is learnt in \emph{end-to-end} fashion during training.

\subsubsection{Context window -- Limits to \ac{LLM} memory}
\label{sec:context}
\acp{LLM} are supposed to conduct long conversations, e.\,g., with mental health patients as in our scenario.
As we saw above, a specific layer inside the standard \ac{LLM} architecture is responsible for connecting each token with each other token inside a sequence.
This allows those models to exploit additional \emph{context} and unlocks a crucial capability of \acp{LLM}, namely that of `keeping in mind' the entire previous conversation, as well as (potentially) prior conversations they have had with the conversation partner/patient in the past.

Ideally, this context would be available in its entirety; in practice, however, it is limited by present computational capabilities, with the main issue being (computer) memory.
In practice, even frontier \acp{LLM} operated by large companies, like Llama-3, place an upper limit to a context of 128,000 tokens~\citep{Dubey24-TL3}.
While computer boundaries are constantly pushed, including additional context is also associated with an increase in operating cost.
What this boils down to is a fundamental limitation of present-day models -- their finite memory.
We discuss ways to overcome this issue in \cref{sec:rag}.

\subsection{Training}
\label{sec:train}
The previous section provided an overview of the inner workings of \acp{LLM}.
In this section, we present the three-step workflow in which they are usually trained. We include a glossary of the most common terms encountered in the training of \acp{LLM} in \cref{box:training-terms}.
%YT: not sure whether this helps, but maybe the following metaphor is helpful: To become a psychotherapist, a fundamental prerequisite is that a human learns how to speak before a special psychotherapist training is possible at all. This is equally true for an LLM and conceptual language learning occurs in a so-called unsupervised pretraining phase (see 3.2.1). However, just knowing the language is not sufficient to be able to conduct psychotherapy. Instead, a human with the capability to speak (metaphorically, the pre-trained LLM) needs to go through a psychotherapist training program, where specific language use is learned by given examples. In the realm of LLM this can be done in a so-called supervised finetuning stage (see 3.2.2). Lastly, even the trained psychotherapist (metaphorically, finetuned LLM) can still improve by feedback from other therapists (e.g., supervision) and hands-on experience with patients. Again, there is a quite similar stage in LLM optimization, in which the LLM learns from human-feedback (see 3.2.3).

However, before diving into the LLM training process, we would like to invite the reader to a small imagination, which helps to conceptually understand the stages outlined below: Imagine someone wants to become a psychotherapist. First, they need to learn the language (e.\,g., by naturally absorbing words and grammar). However, good language qualification is not sufficient for adequate psychotherapy. Instead, one undergoes specialised education, studying therapy techniques through examples and instructions. Finally, even after the specialised training process, one can still refine their skills through experience and supervision. Conceptually, these three steps 1) learning the general language, 2) improving language for a specific task by learning from examples and instructions, and 3) additional improvements by human-feedback are mirrored by the three main stages in LLM training 1) unsupervised pretraining, 2) supervised finetuning, and 3) human reinforcement learning. 
\subsubsection{Unsupervised pretraining}
\acp{LLM} generate \emph{appropriate} responses to the input they receive by computing the most probable next token -- one token at a time.
To do so, they must be trained to provide a good estimate of the following token for any potential combination of previous tokens -- ideally, for all possible previous combinations.
Naturally, given the innumerable combinations that exist for human language, this is practically impossible.
To approximate this, \acp{LLM} are pretrained on enormous corpora of textual data.
They are typically trained to perform the task that will be asked of them at inference -- predict the next token given a sequence of past tokens.
This allows them to estimate the probability distribution for various token combinations.
Yet, while this allows them to form well-structured, `high-probability' sentences, it still results in unconstrained generation which needs to be guided towards the kinds of output that the creators of the model intend it to follow.
Note that this is an optional step for most clinical practitioners (see also \cref{fig:workflows}) but requires substantial technical expertise and compute resources.
In practice, it is recommended to begin from a pretrained \emph{foundation} model~\citep{Bommasani21-OTO}.
%YT: I would think about including a section whether this is needed. IMO I would never recommend to do pre-training on my own except the target/domain language is fundamentally different from normal language (e.g., law). For Psychotherapy I would say, that the domain language is not so different that finetuning is probably sufficient. 

\subsubsection{Supervised finetuning}
%YT: Buzzwords/topics I would include: finetuning with instructions prompts (in context leanring: zero, one, few shot learning); finetuning with examples; Parameter efficient finetuning; reparameterization (e.g., Lora); additive (e.g., prompt tuning with soft prompts)
%AT: I will add this as a box to clarify all this jargon
% \begin{figure}[t]
%     \centering
    
%     \caption{Caption}
%     \label{fig:enter-label}
% \end{figure}
% \begin{pabox}[t]{Title}
%     Test
% \end{pabox}
\begin{figure}[t]
    \centering
    \begin{pabox}[box:training-terms]{Technical terms often encountered when discussing \ac{LLM} training}
    \noindent
    \textbf{PEFT --} stands for \emph{parameter efficient finetuning}, and is a category of methods used for \emph{updating} an \ac{LLM} without changing all of its parameters.
    This is beneficial when lacking the computational resources to adapt the entire model.
    
    \noindent
    \textbf{LoRA --} stands for \emph{low-rank adaptation}, and is a common PEFT method~\citep{Hu21-LoRA}.
    
    \noindent
    \textbf{Prompting --} is the initial input given to the \ac{LLM} by the user or model developer before it begins its generation process.
    An unprompted \ac{LLM} is simply given a `start-of-sequence' token which tells it to begin producing text without any context.
    In case of dual conversations, the user's turn is the prompt fed into the model, potentially augmented with prior turns from both the model and the user.
    In practice, model providers often wrap all inputs to their models in additional prompts that help guide the output and protect against unwanted responses.

    \noindent
    \textbf{Instruction finetuning --} is the process in which an \ac{LLM} is finetuned to respond according to detailed specifications.
    The \emph{instruction} is simply part of the \emph{prompt}.
    Through this type of training, the model learns to recognise instructions that are included in their prompts and respond accordingly.

    \noindent
    \textbf{In-context learning --} is the process in which an \ac{LLM} is taught how to perform a task on-the-fly, i.\,e., \emph{without additional training}~\citep{Dong22-ASO}.
    This can be achieved by supplying a few examples (also known as \emph{few-shot learning}) or describing the task (also known as \emph{zero-shot learning}) as part of the prompt.
    
    \noindent
    \textbf{RLHF --} stands for \emph{reinforcement learning from human feedback}, and is a training paradigm which ranks model responses based on preferences obtained from human annotators~\citep{Bai22-TAH}.
    \end{pabox}
    % \caption{Foo}
    % \label{fig:training-terms}
\end{figure}
%yt: I think the box should be placed somewhere else. currently it is appearing in 3.1. which has nothing to do with training/finetuning. 

After the initial development and pre-training of an LLM its performance can usually be further improved by a supervised finetuning process.  At its core, finetuning is based on data with particular labels.
In other words, counter to the previous stage, the \ac{LLM} has a \emph{target sentence} it is supposed to produce (which it still generates one token at a time).
These target sequences are specified by collecting pairs of inputs and outputs; these encompass anything from general knowledge question-answer pairs, to real human-human conversations that an \ac{LLM} is supposed to mimic, or even curated pairs that are generated by rules and fixed templates (such as a rule to respond to the phrase ``I am thinking of quitting my antidepressants'' by ``Thank you for sharing that with me.'').
This stage of training begins with an \ac{LLM} that can produce coherent language responses that are well-structured, but not necessarily appropriate, and result in an \ac{LLM} conditioned to respond in a particular way when encountering certain inputs.

We note that this stage is often referred to as \emph{instruction finetuning}; however, we consider this to be a subset of input-output pairs that are used for supervised finetuning.
Instructions are more explicit `commands' that control the form that the output \emph{must} take; e.\,g., one can train an \ac{LLM} to ``respond with a simple yes/no'' when given the phrase ``respond with a simple yes/no'' as part of its input.
This allows for \emph{prompting} the \ac{LLM} to produce outputs of a specific form and is often exploited to control \acp{LLM} (sometimes with the user being unaware of it).
We return to this point in \cref{sec:prompt}.
For now, we mention that this supervised finetuning is used to instil specific response patterns as a result of particular inputs, and its scope is much broader than following instructions.

The main assumption here is that there exists a corpus of `correct' input-output pairs for the \ac{LLM} to be trained on.
Oftentimes, such corpora are assembled from datasets containing predictive labels, which are used to train \acp{LLM} in predictive tasks (an aspect of them which we ignore in this contribution).
However, they can include instruction finetuning (see above) but also specific conversation patterns; in the case of psychotherapy, this could be turns from real-life psychotherapy sessions.

Finally, we note here that `learning' in this context remains \emph{statistical} in nature.
While we train \acp{LLM} to faithfully reproduce specific outputs, there is no guarantee -- in fact, no intent either -- that they will exactly reproduce this output when given the same input.
Rather, this step can be seen as an attempt to impart certain \emph{skills} on an \ac{LLM}, such as the ability to respond to a specific input in a specific way, with the hope that it will then \emph{generalise} to similar inputs when put to practice.
This desire for \emph{generalisation} comes at the cost of faithfulness guarantees -- it is a fundamental aspect of \acp{LLM} that they are statistical models; unfortunately, this means that they might occasionally also fail to produce the correct patterns to inputs they have seen in training.
%especially for rare events? Maybe also say that this is the danger in using LLMs for psychotherapy? That rare-symptoms or problems may be misinterpreted or the response may not be matching well? E.g. depression (although it is very heterogenous regarding symptoms) may be matched well, but trichotillomania might "confuse" LLMs?

For example, if a patient with depression were to type the sentence ``I want to stop taking my antidepressants'', an unconditioned \ac{LLM} will respond with a sentence that resembles the text it has seen in its training set; what that might be is thus completely determined by the data used for training.
Therefore, curating a training set that captures the conversations that the \ac{LLM} is expected to encounter during its operating cycle is of fundamental importance.
However, the juxtaposition of the patient's response with a suitable prompt (like ``Answer in a way which appears validating and reassuring'') will result in a response that mostly resembles the combination of prompts/instructions and inputs that has been seen during instruction fine-tuning.
%JL: maybe add an example for similar sentences?: e.g. "find another job"/"get fired"/instead of "having a burnout"? 
%JL: Just a suggestion: again sth out of the psychoworld as example: I am thinking of dropping my antidepressants ``dropping'' and ``antidepressants''?

\subsubsection{Reinforcement learning from human feedback}
Supervised learning conditions \acp{LLM} to respond correctly in cases where the correct response is known.
This is not possible for all potential input sequences.
Especially when it comes to conversations with a human interlocutor, it is not possible to know in advance what all potential conversations will look like -- otherwise, it would have been possible to create a rule-based system in the first place.
Yet, these are exactly the types of conversations that an \ac{LLM} will face in practice, particularly in a psychotherapeutic scenario.

The way to overcome this is to directly collect human feedback on the appropriateness of a specific response \emph{without knowing what the exact response should be}.
This shortcut is achieved using reinforcement learning, with the \emph{reward} that \emph{reinforces} a particular type of responses being human ratings collected over a large set of \ac{LLM} responses.
This allows practitioners to further finetune an \ac{LLM} (as a third step after pretraining and supervised finetuning) in a way that maximises this reward -- a process reminiscent of operant conditioning in \ac{CBT}~\citep{staddon2003operant}.
%JL:fun fact: reinforcement learning is basically the ground truth of cognitive behavioural therapy (=operant conditioning)

The success of this paradigm hinges on the crowd of annotators used to rate responses.
Given how costly this step might be, it is not uncommon for data to be collected using residents of low-income countries. 
%BS: You already say low-income countries - it is irrelevant here, if this is from Africa, but a reference will still be good.
% (a report tracked most annotators used for this step in ChatGPT to African countries\todo{cite}).
In the case of psychotherapy, it is highly unlikely that existing datasets -- and thus models -- have been annotated by mental health patients or health care providers. 
This calls into question the reliability of these models for our target scenario, especially since the effect of an intervention (delivered in the form of an \ac{LLM} response) can be very subjective in nature, and requires feedback from the patients themselves.
Translating that to clinical practice, where feedback must come from patients already burdened with a heavy mental load, is challenging, but can substantially improve the reliability and applicability of \acp{LLM} in clinical practice.
To that end, we may envision the integration of patient groups as volunteers contributing feedback that helps steer \acp{LLM} towards responses that are more appropriate for their interest groups -- similar to the community efforts driving voluntary participation in clinical trials.
% It further challenges the utility of this technique in a psychotherapy setup, as it is difficult to envision a future where mental health patients will voluntarily annotate the responses of a model in the quantity needed to apply this technique in practice, given their already high mental burden.
% Thus, we reference it mainly for the sake of completeness in covering how modern \acp{LLM} are trained.
%JL: why? because their mental burden is too high? Maybe they would be interested in helping out? Especially so called "lived experience experts" or with bad therapy experiences might be very motivated in doing so...
%JC - I suppose one challenge would be its hard to distinguish between subjective vs objective assessments (e.g. patient perceived helpfulness of response may be different from objective helpfulness)? But I agree with the above that we might not want to rule out voluntary feedback. In fact, this might be something we want to work towards right? Not necessarily RLHF, but ways to collect patient input for e.g. personalisation

%YT: What about consitutional AI
%YT: What about a short introduciton into evaluation metrics. 

\subsection{Controlling the generation}
\label{sec:control}
We have already touched upon the issue of controllability in the previous section.
Given the sensitive nature of psychotherapy, it is natural to want a way to establish specific \emph{guardrails} around the output of an \ac{LLM}.
Moreover, given the very specialised knowledge required to handle mental health patients, another requirement is the ability to inject (new) knowledge to a model after training.
Finally, given the personalised nature of psychotherapy conversations, it is vital that the \ac{LLM} maintains the patient's history `in-mind' when producing each output.
Those requirements are presently tackled through the use of \emph{prompting} and its more advanced `sister' technique, \ac{RAG}.
Though far from perfect, these techniques can partially mitigate controllability and memory issues, though further research is required to completely alleviate them.
%JC - To me this reads like we're claiming that prompt engineering can already fully satisfy these safety, knowledge injection, and user history incorporation requirements. Can we rephrase/add a short disclaimer here? (e.g. present prompting as a part of a(n imperfect) toolbox that we want to responsibly apply to psychotherapy)

\subsubsection{Prompting}
\label{sec:prompt}
Prompting is the process of extending an \ac{LLM}'s input to incorporate additional information or \emph{instructions} about how it should perform.
This helps to `set the scene' for an \ac{LLM} as exhibited by widely-circulating prompts instructing an \ac{LLM} to ``respond in a friendly tone'' or similar.
Additionally, they can be used to integrate background information about the patient, e.\,g., by co-opting results from standardised evaluation metrics and demographics (``this is a female college student who recently scored 20 on the PHQ-9 scale''). %that's really good!
Prompts are model- and context-specific; this means that the success of a prompt can only be measured in a trial-and-fail process that involves conversations with actual patients (though an initial search can be done in advance).
Moreover, prompts can be sensitive to formatting, and only partially alleviate the problem of controllability, given that the nature of \acp{LLM} remains stochastic -- meaning that even given the same input, a model will still generate a different output~\citep{Sclar24-QLM}.
%JC - consider mentioning prompt sensitivity to formatting + emphasising stochastic nature of LLM? For example see Sclar et al 2024 https://arxiv.org/abs/2310.11324
While this risk of failure is a major concern for the sensitive nature of psychotherapy, the potential benefits from prompting can be substantial; how it can be employed in practice remains to be seen.

\subsubsection{Retrieval augmented generation}
%YT: Is this needed? If so, what about general information about deployment (e.g., distillation, quantization, pruning), why not other augmentation methods besides RAG (e.g., chain-of-thought, program-aided language models, ReAct, LangChain
\label{sec:rag}
In the previous subsection, we described \emph{manual} prompts that add additional context or constrain the output.
The same underlying mechanism can be utilised for \emph{automatic} prompts.
These serve to incorporate new information that was not available to the \ac{LLM} during training as part of a technique (or a family of techniques) named \acf{RAG}.  %BS: RAG was already introduced as abbreviation above - is this intentionally re-introduced?
\ac{RAG} relies on searching a database of textual documents (also referred to as \emph{assets}) for information related to a \emph{query}; in the simplest case, this query is the initial input to the \ac{LLM} (i.\,e., before it starts generating).

Interestingly, this same technique can be used to artificially enhance the memory of an \ac{LLM} by storing relevant information for a patient (e.\,g., collected in prior sessions) and retrieving them when it becomes relevant during the current session.
Given that the context for most \acp{LLM} reaches up to 200,000 tokens, a substantial part of the load can be offloaded to \ac{RAG} for recovering relevant snippets of past conversations (even from the current session if its length is larger than the context window).
The main challenge lies with identifying which parts of a conversation are relevant for storing in the database that needs to be queried without knowing what the query will be in advance (of course, everything can be stored, but then the retrieval will be very computationally demanding); however, this part can be automated using language analysis as well (either by relying on the same \ac{LLM}, or using another, specialised one).

%YT: Other topics maybe worth mentioning. toxicity (NO GO for psychotherapy LLM), Hallucinaiotion, intellectual property; LLM's Newton Law aka emergent LLM capabilites

\subsection{Limitations}
% JC - I have collated YT + JL + my previous comment into this draft limitation section. Please edit as you see fit.
% Perhaps not all of these points are in-scope for this paper, and I am wary of overstating the limitations of LLMs. Nonetheless I think these are issues the psychologist should know about; this way they can ask the right questions when comparing which LMs they want to try in their practice. Similarly developers should communicate why and how they have/havent't addressed these if "selling" or sharing their tools for clinical applications. Even if they don't "solve" these issues they should acknowledge they exist and design the systems for psychologists to adopt with these in mind. 
While \acp{LLM} show considerable promise, they possess technical and ethical limitations, including the following.
Importantly, translating them into clinical practice requires a thorough and critical examination of their strenghts and limitations.
Once again, this calls for an integrative approach to bridge the divide across communities, with psychologists on the one hand inquiring about known limitations, and model developers on the other being transparent about the risks related to their models.

\paragraph{Hallucinations}
\acp{LLM} can generate plausible-sounding but factually incorrect information~\citep{Huang2023ASO}. This could lead to potentially harmful advice, misrepresentation of therapeutic techniques, or inaccurate clinical information.

\paragraph{Biases} \acp{LLM} may amplify existing prejudices, leading to low quality or harmful responses~\citep{gabriel2024airelatetestinglarge} and differential treatment recommendations~\citep{omiye2023large}. While there are bias mitigation techniques~\citep{gallegos2024bias}, they should not over-correct to harm clinical utility (e.\,g., models should be fair and still respond to real between-group psychopathological differences).

\paragraph{Privacy}
\acp{LLM} can memorise and inadvertently disclose information from their training data~\citep{carlini2019secret}. They also struggle to reason about what information is shareable in which contexts~\citep{mireshghallah2024can}, which may compromise confidentiality. These risks necessitate strict data protection measures when integrating \acp{LLM} into workflows and documentation systems. 

\paragraph{Syncopathy}
\acp{LLM} trained on human feedback may prioritise user agreement over response truthfulness~\citep{sharma2024towards}. This could manifest as the model avoiding therapeutic confrontation or inappropriately validating destructive behaviours.

\paragraph{Safety}
\acp{LLM} may generate harmful content or encourage user overdependence~\citep{ma2023understanding}. They can fail to respond appropriately to distress~\citep{heston2023safety} and to escalate scenarios to human professionals when needed~\citep{de2024chatbots}. These risks underscore the need to design systems with rigorous safeguards and clear pathways to human supervision and intervention.

Understanding these limitations is crucial: Psychologists need to evaluate tools and provide informed feedback, while developers must communicate their systems' intended use, shortcomings, and management strategies, even when complete solutions are not yet feasible. This shared understanding forms the foundation for responsible and effective \ac{LLM} implementation in psychotherapy.

\section{Conclusion}
We presented a distilled, easily-digestible overview of how \ac{LLM} building blocks, training scheme, and usage techniques coalesce into their phenomenal success as conversational engines that are taking the world by storm.
In a nutshell, \acp{LLM} are stochastic, autoregressive models of language -- given a text input, or \emph{prompt}, they will produce a text output which serves as a good \emph{continuation} of the input.
This output will be unique each time and is generated one token at at time.
The generation process leverages patterns learnt from data during training and accounts for the input context to generate coherent responses.
As such, it can be used to support numerous functions relevant for clinical practice of psychologists, from note summarisation to running autonomous psychotherapeutic sessions.
Given present barriers to treatment, we expect this technology to play a pivotal role in an improved delivery of care for mental disorders.
Crucially, their responsible development and translation to clinical practice entails the closer collaboration between the psychology and technical communities.
We hope that this overview will lead to increased involvement of the psychology community in future development of \acp{LLM} and lead to a tighter future collaboration between the two communities.

\section*{Acknowledgments}
This work has received funding from the DFG’s Reinhart Koselleck project No.~442218748 (AUDI0NOMOUS).
% Moreover, ...

%Bibliography
\section{\refname}
\printbibliography[heading=none] 

@article{Weizenbaum66-EAC,
  title={ELIZA—a computer program for the study of natural language communication between man and machine},
  author={Weizenbaum, Joseph},
  journal={Communications of the ACM},
  volume={9},
  number={1},
  pages={36--45},
  year={1966},
  publisher={ACM New York, NY, USA}
}

@article{seiferth2023mental,
  title={How to e-mental health: a guideline for researchers and practitioners using digital technology in the context of mental health},
  author={Seiferth, Caroline and Vogel, Lea and Aas, Benjamin and Brandhorst, Isabel and Carlbring, Per and Conzelmann, Annette and Esfandiari, Narges and Finkbeiner, Marlene and Hollmann, Karsten and Lautenbacher, Heinrich and others},
  journal={Nature mental health},
  volume={1},
  number={8},
  pages={542--554},
  year={2023},
  publisher={Nature Publishing Group US New York}
}

@article{cuijpers2019effectiveness,
  title={Effectiveness and acceptability of cognitive behavior therapy delivery formats in adults with depression: a network meta-analysis},
  author={Cuijpers, Pim and Noma, Hisashi and Karyotaki, Eirini and Cipriani, Andrea and Furukawa, Toshi A},
  journal={JAMA psychiatry},
  volume={76},
  number={7},
  pages={700--707},
  year={2019},
  publisher={American Medical Association}
}

@article{hedman2023therapist,
  title={Therapist-supported Internet-based cognitive behaviour therapy yields similar effects as face-to-face therapy for psychiatric and somatic disorders: an updated systematic review and meta-analysis},
  author={Hedman-Lagerl{\"o}f, Erik and Carlbring, Per and Sv{\"a}rdman, Frank and Riper, Heleen and Cuijpers, Pim and Andersson, Gerhard},
  journal={World Psychiatry},
  volume={22},
  number={2},
  pages={305--314},
  year={2023},
  publisher={Wiley Online Library}
}

@article{vaidyam2021changes,
  title={Changes to the Psychiatric Chatbot Landscape: A Systematic Review of Conversational Agents in Serious Mental Illness: Changements du paysage psychiatrique des chatbots: une revue syst{\'e}matique des agents conversationnels dans la maladie mentale s{\'e}rieuse},
  author={Vaidyam, Aditya Nrusimha and Linggonegoro, Danny and Torous, John},
  journal={The Canadian Journal of Psychiatry},
  volume={66},
  number={4},
  pages={339--348},
  year={2021},
  publisher={Sage Publications Sage CA: Los Angeles, CA}
}

@article{Elhage21-AMF,
  title={A mathematical framework for transformer circuits},
  author={Elhage, Nelson and Nanda, Neel and Olsson, Catherine and Henighan, Tom and Joseph, Nicholas and Mann, Ben and Askell, Amanda and Bai, Yuntao and Chen, Anna and Conerly, Tom and others},
  journal={White Paper},
  volume={1},
  number={1},
  pages={12},
  year={2021}
}

@article{Dong22-ASO,
  title={A survey on in-context learning},
  author={Dong, Qingxiu and Li, Lei and Dai, Damai and Zheng, Ce and Wu, Zhiyong and Chang, Baobao and Sun, Xu and Xu, Jingjing and Sui, Zhifang},
  journal={arXiv preprint arXiv:2301.00234},
  year={2022}
}

@article{Bai22-TAH,
  title={Training a helpful and harmless assistant with reinforcement learning from human feedback},
  author={Bai, Yuntao and Jones, Andy and Ndousse, Kamal and Askell, Amanda and Chen, Anna and DasSarma, Nova and Drain, Dawn and Fort, Stanislav and Ganguli, Deep and Henighan, Tom and others},
  journal={arXiv preprint arXiv:2204.05862},
  year={2022}
}

@inproceedings{Hu21-LoRA,
  title={LoRA: Low-Rank Adaptation of Large Language Models},
  author={Hu, Edward J and Wallis, Phillip and Allen-Zhu, Zeyuan and Li, Yuanzhi and Wang, Shean and Wang, Lu and Chen, Weizhu and others},
  booktitle={Proceedings of the International Conference on Learning Representations},
  year={2021}
}

@article{monteith2022expectations,
  title={Expectations for artificial intelligence (AI) in psychiatry},
  author={Monteith, Scott and Glenn, Tasha and Geddes, John and Whybrow, Peter C and Achtyes, Eric and Bauer, Michael},
  journal={Current Psychiatry Reports},
  volume={24},
  number={11},
  pages={709--721},
  year={2022},
  publisher={Springer}
}

@article{VanDaele2022online,
   author = {Tom Van Daele and Kim Mathiasen and Per Carlbring and Sylvie Bernaerts and Agostino Brugnera and Angelo Compare and Aranzazu Duque and Jonas Eimontas and David Gosar and Lise Haddouk and Maria Karekla and Pia Larsen and Gianluca Lo Coco and Tine Nordgreen and João Salgado and Andreas R. Schwerdtfeger and Eva Van Assche and Sam Willems and Nele A.J. De Witte},
   doi = {10.1016/j.invent.2022.100571},
   issn = {22147829},
   journal = {Internet Interventions},
   month = {12},
   pages = {100571},
   title = {Online consultations in mental healthcare: Modelling determinants of use and experience based on an international survey study at the onset of the pandemic},
   volume = {30},
   year = {2022},
}

@article{Boyd22-TDA,
  title={The development and psychometric properties of LIWC-22},
  author={Boyd, Ryan L and Ashokkumar, Ashwini and Seraj, Sarah and Pennebaker, James W},
  journal={Austin, TX: University of Texas at Austin},
  volume={10},
  year={2022}
}

@article{Demszky23-ULL,
  title={Using large language models in psychology},
  author={Demszky, Dorottya and Yang, Diyi and Yeager, David S and Bryan, Christopher J and Clapper, Margarett and Chandhok, Susannah and Eichstaedt, Johannes C and Hecht, Cameron and Jamieson, Jeremy and Johnson, Meghann and others},
  journal={Nature Reviews Psychology},
  volume={2},
  number={11},
  pages={688--701},
  year={2023},
  publisher={Nature Publishing Group US New York}
}

@article{baumel2019there,
  title={Is there a trial bias impacting user engagement with unguided e-mental health interventions? A systematic comparison of published reports and real-world usage of the same programs},
  author={Baumel, Amit and Edan, Stav and Kane, John M},
  journal={Translational behavioral medicine},
  volume={9},
  number={6},
  pages={1020--1033},
  year={2019},
  publisher={Oxford University Press US}
}

@article{moshe2021digital,
  title={Digital interventions for the treatment of depression: A meta-analytic review.},
  author={Moshe, Isaac and Terhorst, Yannik and Philippi, Paula and Domhardt, Matthias and Cuijpers, Pim and Cristea, Ioana and Pulkki-R{\aa}back, Laura and Baumeister, Harald and Sander, Lasse B},
  journal={Psychological bulletin},
  volume={147},
  number={8},
  pages={749},
  year={2021},
  publisher={American Psychological Association}
}

@article{Philippi21-ATD,
  title={Acceptance towards digital health interventions--model validation and further development of the unified theory of acceptance and use of technology},
  author={Philippi, Paula and Baumeister, Harald and Apolin{\'a}rio-Hagen, Jennifer and Ebert, David Daniel and Hennemann, Severin and Kott, Leonie and Lin, Jiaxi and Messner, Eva-Maria and Terhorst, Yannik},
  journal={Internet interventions},
  volume={26},
  pages={100459},
  year={2021},
  publisher={Elsevier}
}

@inproceedings{Sclar24-QLM,
  title={Quantifying Language Models' Sensitivity to Spurious Features in Prompt Design or: How I learned to start worrying about prompt formatting},
  author={Sclar, Melanie and Choi, Yejin and Tsvetkov, Yulia and Suhr, Alane},
  booktitle={Proceedings of the International Conference on Learning Representations},
  year={2024}
}

@article{staddon2003operant,
  title={Operant conditioning},
  author={Staddon, John ER and Cerutti, Daniel T},
  journal={Annual review of psychology},
  volume={54},
  number={1},
  pages={115--144},
  year={2003},
  publisher={Annual Reviews 4139 El Camino Way, PO Box 10139, Palo Alto, CA 94303-0139, USA}
}

@article{Li2023,
   author = {Han Li and Renwen Zhang and Yi-Chieh Lee and Robert E. Kraut and David C. Mohr},
   doi = {10.1038/s41746-023-00979-5},
   issn = {2398-6352},
   issue = {1},
   journal = {npj Digital Medicine},
   month = {12},
   pages = {236},
   title = {Systematic review and meta-analysis of AI-based conversational agents for promoting mental health and well-being},
   volume = {6},
   year = {2023},
}

@article{abd2019overview,
  title={An overview of the features of chatbots in mental health: A scoping review},
  author={Abd-Alrazaq, Alaa A and Alajlani, Mohannad and Alalwan, Ali Abdallah and Bewick, Bridgette M and Gardner, Peter and Househ, Mowafa},
  journal={International journal of medical informatics},
  volume={132},
  pages={103978},
  year={2019},
  publisher={Elsevier}
}

@article{seizer2024primer,
  title={A primer on sampling rates of ambulatory assessments.},
  author={Seizer, Lennart and Schiepek, G{\"u}nter and Cornelissen, Germaine and L{\"o}chner, Johanna},
  journal={Psychological Methods},
  year={2024},
  publisher={American Psychological Association}
}

@article{chang2018electronic,
  title={Electronic media exposure and use among toddlers},
  author={Chang, Hyoung Yoon and Park, Eun-Jin and Yoo, Hee-Jeong and won Lee, Jee and Shin, Yunmi},
  journal={Psychiatry investigation},
  volume={15},
  number={6},
  pages={568},
  year={2018},
  publisher={Korean Neuropsychiatric Association}
}

@article{Buck2024,
   author = {Benjamin Buck and Arya Kadakia and Anna Larsen and Justin Tauscher and Jessy Guler and Dror Ben-Zeev},
   doi = {10.1037/pri0000250},
   issn = {2377-8903},
   journal = {Practice Innovations},
   month = {7},
   title = {Digital interventions for people waitlisted for mental health services: A needs assessment and preference survey.},
   year = {2024},
}

@article{Vaswani17-AIA,
  title={Attention is all you need},
  author={Vaswani, Ashish and Shazeer, Noam and Parmar, Niki and Uszkoreit, Jakob and Jones, Llion and Gomez, Aidan N and Kaiser, {\L}ukasz and Polosukhin, Illia},
  journal={Advances in neural information processing systems},
  volume={30},
  year={2017}
}

@techreport{WorldHealthOrganization2022,
address = {Geneva},
author = {{World Health Organization}},
isbn = {978-92-4-004933-8},
title = {{World mental health report: transforming mental health for all}},
url = {https://www.who.int/publications/i/item/9789240049338},
year = {2022}
}

@article{Herrman2022,
author = {Herrman, Helen and Patel, Vikram and Kieling, Christian and Berk, Michael and Buchweitz, Claudia and Cuijpers, Pim and Furukawa, Toshiaki A. and Kessler, Ronald C. and Kohrt, Brandon A. and Maj, Mario and McGorry, Patrick and Reynolds, Charles F. and Weissman, Myrna M. and Chibanda, Dixon and Dowrick, Christopher and Howard, Louise M. and Hoven, Christina W. and Knapp, Martin and Mayberg, Helen S. and Penninx, Brenda W J H and Xiao, Shuiyuan and Trivedi, Madhukar and Uher, Rudolf and Vijayakumar, Lakshmi and Wolpert, Miranda},
doi = {10.1016/S0140-6736(21)02141-3},
issn = {01406736},
journal = {The Lancet},
month = {mar},
number = {10328},
pages = {957--1022},
pmid = {35180424},
title = {{Time for united action on depression: a Lancet–World Psychiatric Association Commission}},
url = {https://linkinghub.elsevier.com/retrieve/pii/S0140673621021413},
volume = {399},
year = {2022}
}

@article{Moshe2021,
author = {Moshe, Isaac and Terhorst, Yannik and Philippi, Paula and Domhardt, Matthias and Cuijpers, Pim and Cristea, Ioana and Pulkki-R{\aa}back, Laura and Baumeister, Harald and Sander, Lasse B.},
doi = {10.1037/bul0000334},
issn = {1939-1455},
journal = {Psychological Bulletin},
month = {aug},
number = {8},
pages = {749--786},
title = {{Digital interventions for the treatment of depression: A meta-analytic review.}},
url = {http://doi.apa.org/getdoi.cfm?doi=10.1037/bul0000334 https://doi.apa.org/doi/10.1037/bul0000334},
volume = {147},
year = {2021}
}

@article{Terhorst2024,
author = {Terhorst, Yannik and Kaiser, Tim and Brakemeier, Eva-Lotta and Moshe, Isaac and Philippi, Paula and Cuijpers, Pim and Baumeister, Harald and Sander, Lasse Bosse},
doi = {10.1001/jamanetworkopen.2024.23241},
issn = {2574-3805},
journal = {JAMA Network Open},
month = {jul},
number = {7},
pages = {e2423241},
title = {{Heterogeneity of Treatment Effects in Internet- and Mobile-Based Interventions for Depression}},
url = {https://jamanetwork.com/journals/jamanetworkopen/fullarticle/2821335},
volume = {7},
year = {2024}
}

@article{Qiu24-IAS,
  title={Interactive Agents: Simulating Counselor-Client Psychological Counseling via Role-Playing LLM-to-LLM Interactions},
  author={Qiu, Huachuan and Lan, Zhenzhong},
  journal={arXiv preprint arXiv:2408.15787},
  year={2024}
}

@inproceedings{Kundan21-GSN,
  title={Generating SOAP Notes from Doctor-Patient Conversations Using Modular Summarization Techniques},
  author={Krishna, Kundan and Khosla, Sopan and Bigham, Jeffrey P and Lipton, Zachary C},
  booktitle={Proceedings of the 59th Annual Meeting of the Association for Computational Linguistics and the 11th International Joint Conference on Natural Language Processing (Volume 1: Long Papers)},
  pages={4958--4972},
  year={2021}
}

@article{Bendig22-TNG,
  title={The next generation: chatbots in clinical psychology and psychotherapy to foster mental health--a scoping review},
  author={Bendig, Eileen and Erb, Benjamin and Schulze-Thuesing, Lea and Baumeister, Harald},
  journal={Verhaltenstherapie},
  volume={32},
  number={Suppl. 1},
  pages={64--76},
  year={2022},
  publisher={S. Karger AG}
}

@inproceedings{Vijayakumar18-DBS, 
    title={Diverse Beam Search for Improved Description of Complex Scenes}, 
    volume={32},
    number={1},
    journal={Proceedings of the AAAI Conference on Artificial Intelligence},
    author={Vijayakumar, Ashwin and Cogswell, Michael and Selvaraju, Ramprasaath and Sun, Qing and Lee, Stefan and Crandall, David and Batra, Dhruv},
    year={2018}
}

@article{Mikolov13-DRO,
  title={Distributed representations of words and phrases and their compositionality},
  author={Mikolov, Tomas and Sutskever, Ilya and Chen, Kai and Corrado, Greg S and Dean, Jeff},
  journal={Advances in neural information processing systems},
  volume={26},
  year={2013}
}

@inproceedings{Pennington14-GGV,
  title={Glove: Global vectors for word representation},
  author={Pennington, Jeffrey and Socher, Richard and Manning, Christopher D},
  booktitle={Proceedings of the 2014 conference on empirical methods in natural language processing (EMNLP)},
  pages={1532--1543},
  year={2014}
}

@inproceedings{Jean15-OUV,
  title={On Using Very Large Target Vocabulary for Neural Machine Translation},
  author={Jean, S{\'e}bastien and Cho, Kyunghyun and Memisevic, Roland and Bengio, Yoshua},
  booktitle={Proceedings of the 53rd Annual Meeting of the Association for Computational Linguistics and the 7th International Joint Conference on Natural Language Processing (Volume 1: Long Papers)},
  pages={1--10},
  year={2015}
}

@article{Dubey24-TL3,
  title={The llama 3 herd of models},
  author={Dubey, Abhimanyu and Jauhri, Abhinav and Pandey, Abhinav and Kadian, Abhishek and Al-Dahle, Ahmad and Letman, Aiesha and Mathur, Akhil and Schelten, Alan and Yang, Amy and Fan, Angela and others},
  journal={arXiv preprint arXiv:2407.21783},
  year={2024}
}

@article{Sennrich16-NMT,
  title={Neural machine translation of rare words with subword units},
  author={Sennrich, Rico and Haddow, B, and Birch, A},
  journal={Proceedings of the 54th Annual Meeting of the Association for Computational Linguistics.},
  year={2016}
}

@article{He23-CAI,
  title={Conversational agent interventions for mental health problems: systematic review and meta-analysis of randomized controlled trials},
  author={He, Yuhao and Yang, Li and Qian, Chunlian and Li, Tong and Su, Zhengyuan and Zhang, Qiang and Hou, Xiangqing},
  journal={Journal of medical Internet research},
  volume={25},
  pages={e43862},
  year={2023},
  publisher={JMIR Publications Toronto, Canada}
}

@inproceedings{Bender20-CTN,
  title={Climbing towards NLU: On meaning, form, and understanding in the age of data},
  author={Bender, Emily M and Koller, Alexander},
  booktitle={Proceedings of the 58th annual meeting of the association for computational linguistics},
  pages={5185--5198},
  year={2020}
}

@article{Bommasani21-OTO,
  title={On the opportunities and risks of foundation models},
  author={Bommasani, Rishi and Hudson, Drew A and Adeli, Ehsan and Altman, Russ and Arora, Simran and von Arx, Sydney and Bernstein, Michael S and Bohg, Jeannette and Bosselut, Antoine and Brunskill, Emma and others},
  journal={arXiv preprint arXiv:2108.07258},
  year={2021}
}

@article{James2018,
author = {James, Spencer L. and Abate, Degu and Abate, Kalkidan Hassen and Abay, Solomon M. and Abbafati, Cristiana and Abbasi, Nooshin and Abbastabar, Hedayat and Abd-Allah, Foad and Abdela, Jemal and Abdelalim, Ahmed and Abdollahpour, Ibrahim and Abdulkader, Rizwan Suliankatchi and Abebe, Zegeye and Abera, Semaw F. and Abil, Olifan Zewdie and Abraha, Haftom Niguse and Abu-Raddad, Laith Jamal and Abu-Rmeileh, Niveen M E and Accrombessi, Manfred Mario Kokou and Acharya, Dilaram and Acharya, Pawan and Ackerman, Ilana N. and Adamu, Abdu A. and Adebayo, Oladimeji M. and Adekanmbi, Victor and Adetokunboh, Olatunji O. and Adib, Mina G. and Adsuar, Jose C. and Afanvi, Kossivi Agbelenko and Afarideh, Mohsen and Afshin, Ashkan and Agarwal, Gina and Agesa, Kareha M. and Aggarwal, Rakesh and Aghayan, Sargis Aghasi and Agrawal, Sutapa and Ahmadi, Alireza and Ahmadi, Mehdi and Ahmadieh, Hamid and Ahmed, Muktar Beshir and Aichour, Amani Nidhal and Aichour, Ibtihel and Aichour, Miloud Taki Eddine and Akinyemiju, Tomi and Akseer, Nadia and Al-Aly, Ziyad and Al-Eyadhy, Ayman and Al-Mekhlafi, Hesham M. and Al-Raddadi, Rajaa M. and Alahdab, Fares and Alam, Khurshid and Alam, Tahiya and Alashi, Alaa and Alavian, Seyed Moayed and Alene, Kefyalew Addis and Alijanzadeh, Mehran and Alizadeh-Navaei, Reza and Aljunid, Syed Mohamed and Alkerwi, Ala'a and Alla, Fran{\c{c}}ois and Allebeck, Peter and Alouani, Mohamed M L and Altirkawi, Khalid and Alvis-Guzman, Nelson and Amare, Azmeraw T. and Aminde, Leopold N. and Ammar, Walid and Amoako, Yaw Ampem and Anber, Nahla Hamed and Andrei, Catalina Liliana and Androudi, Sofia and Animut, Megbaru Debalkie and Anjomshoa, Mina and Ansha, Mustafa Geleto and Antonio, Carl Abelardo T. and Anwari, Palwasha and Arabloo, Jalal and Arauz, Antonio and Aremu, Olatunde and Ariani, Filippo and Armoon, Bahroom and {\"{A}}rnl{\"{o}}v, Johan and Arora, Amit and Artaman, Al and Aryal, Krishna K. and Asayesh, Hamid and Asghar, Rana Jawad and Ataro, Zerihun and Atre, Sachin R. and Ausloos, Marcel and Avila-Burgos, Leticia and Avokpaho, Euripide F G A and Awasthi, Ashish and {Ayala Quintanilla}, Beatriz Paulina and Ayer, Rakesh and Azzopardi, Peter S. and Babazadeh, Arefeh and Badali, Hamid and Badawi, Alaa and Bali, Ayele Geleto and Ballesteros, Katherine E. and Ballew, Shoshana H. and Banach, Maciej and Banoub, Joseph Adel Mattar and Banstola, Amrit and Barac, Aleksandra and Barboza, Miguel A. and Barker-Collo, Suzanne Lyn and B{\"{a}}rnighausen, Till Winfried and Barrero, Lope H. and Baune, Bernhard T. and Bazargan-Hejazi, Shahrzad and Bedi, Neeraj and Beghi, Ettore and Behzadifar, Masoud and Behzadifar, Meysam and B{\'{e}}jot, Yannick and Belachew, Abate Bekele and Belay, Yihalem Abebe and Bell, Michelle L. and Bello, Aminu K. and Bensenor, Isabela M. and Bernabe, Eduardo and Bernstein, Robert S. and Beuran, Mircea and Beyranvand, Tina and Bhala, Neeraj and Bhattarai, Suraj and Bhaumik, Soumyadeep and Bhutta, Zulfiqar A. and Biadgo, Belete and Bijani, Ali and Bikbov, Boris and Bilano, Ver and Bililign, Nigus and {Bin Sayeed}, Muhammad Shahdaat and Bisanzio, Donal and Blacker, Brigette F. and Blyth, Fiona M. and Bou-Orm, Ibrahim R. and Boufous, Soufiane and Bourne, Rupert and Brady, Oliver J. and Brainin, Michael and Brant, Luisa C. and Brazinova, Alexandra and Breitborde, Nicholas J K and Brenner, Hermann and Briant, Paul Svitil and Briggs, Andrew M. and Briko, Andrey Nikolaevich and Britton, Gabrielle and Brugha, Traolach and Buchbinder, Rachelle and Busse, Reinhard and Butt, Zahid A. and Cahuana-Hurtado, Lucero and Cano, Jorge and C{\'{a}}rdenas, Rosario and Carrero, Juan J. and Carter, Austin and Carvalho, F{\'{e}}lix and Casta{\~{n}}eda-Orjuela, Carlos A. and {Castillo Rivas}, Jacqueline and Castro, Franz and Catal{\'{a}}-L{\'{o}}pez, Ferr{\'{a}}n and Cercy, Kelly M. and Cerin, Ester and Chaiah, Yazan and Chang, Alex R and Chang, Hsing-Yi and Chang, Jung-Chen and Charlson, Fiona J. and Chattopadhyay, Aparajita and Chattu, Vijay Kumar and Chaturvedi, Pankaj and Chiang, Peggy Pei-Chia and Chin, Ken Lee and Chitheer, Abdulaal and Choi, Jee-Young J and Chowdhury, Rajiv and Christensen, Hanne and Christopher, Devasahayam J. and Cicuttini, Flavia M. and Ciobanu, Liliana G. and Cirillo, Massimo and Claro, Rafael M. and Collado-Mateo, Daniel and Cooper, Cyrus and Coresh, Josef and Cortesi, Paolo Angelo and Cortinovis, Monica and Costa, Megan and Cousin, Ewerton and Criqui, Michael H. and Cromwell, Elizabeth A. and Cross, Marita and Crump, John A. and Dadi, Abel Fekadu and Dandona, Lalit and Dandona, Rakhi and Dargan, Paul I. and Daryani, Ahmad and {Das Gupta}, Rajat and {Das Neves}, Jos{\'{e}} and Dasa, Tamirat Tesfaye and Davey, Gail and Davis, Adrian C. and Davitoiu, Dragos Virgil and {De Courten}, Barbora and {De La Hoz}, Fernando Pio and {De Leo}, Diego and {De Neve}, Jan-Walter and Degefa, Meaza Girma and Degenhardt, Louisa and Deiparine, Selina and Dellavalle, Robert P. and Demoz, Gebre Teklemariam and Deribe, Kebede and Dervenis, Nikolaos and {Des Jarlais}, Don C. and Dessie, Getenet Ayalew and Dey, Subhojit and Dharmaratne, Samath Dhamminda and Dinberu, Mesfin Tadese and Dirac, M. Ashworth and Djalalinia, Shirin and Doan, Linh and Dokova, Klara and Doku, David Teye and Dorsey, E. Ray and Doyle, Kerrie E. and Driscoll, Tim Robert and Dubey, Manisha and Dubljanin, Eleonora and Duken, Eyasu Ejeta and Duncan, Bruce B. and Duraes, Andre R. and Ebrahimi, Hedyeh and Ebrahimpour, Soheil and Echko, Michelle Marie and Edvardsson, David and Effiong, Andem and Ehrlich, Joshua R. and {El Bcheraoui}, Charbel and {El Sayed Zaki}, Maysaa and El-Khatib, Ziad and Elkout, Hajer and Elyazar, Iqbal R F and Enayati, Ahmadali and Endries, Aman Yesuf and Er, Benjamin and Erskine, Holly E. and Eshrati, Babak and Eskandarieh, Sharareh and Esteghamati, Alireza and Esteghamati, Sadaf and Fakhim, Hamed and {Fallah Omrani}, Vahid and Faramarzi, Mahbobeh and Fareed, Mohammad and Farhadi, Farzaneh and Farid, Talha A. and s{\'{a}} Farinha, Carla Sofia E and Farioli, Andrea and Faro, Andre and Farvid, Maryam S. and Farzadfar, Farshad and Feigin, Valery L. and Fentahun, Netsanet and Fereshtehnejad, Seyed-Mohammad and Fernandes, Eduarda and Fernandes, Joao C. and Ferrari, Alize J. and Feyissa, Garumma Tolu and Filip, Irina and Fischer, Florian and Fitzmaurice, Christina and Foigt, Nataliya A. and Foreman, Kyle J. and Fox, Jack and Frank, Tahvi D. and Fukumoto, Takeshi and Fullman, Nancy and F{\"{u}}rst, Thomas and Furtado, Jo{\~{a}}o M. and Futran, Neal D. and Gall, Seana and Ganji, Morsaleh and Gankpe, Fortune Gbetoho and Garcia-Basteiro, Alberto L. and Gardner, William M. and Gebre, Abadi Kahsu and Gebremedhin, Amanuel Tesfay and Gebremichael, Teklu Gebrehiwo and Gelano, Tilayie Feto and Geleijnse, Johanna M. and Genova-Maleras, Ricard and Geramo, Yilma Chisha Dea and Gething, Peter W. and Gezae, Kebede Embaye and Ghadiri, Keyghobad and {Ghasemi Falavarjani}, Khalil and Ghasemi-Kasman, Maryam and Ghimire, Mamata and Ghosh, Rakesh and Ghoshal, Aloke Gopal and Giampaoli, Simona and Gill, Paramjit Singh and Gill, Tiffany K. and Ginawi, Ibrahim Abdelmageed and Giussani, Giorgia and Gnedovskaya, Elena V. and Goldberg, Ellen M. and Goli, Srinivas and G{\'{o}}mez-Dant{\'{e}}s, Hector and Gona, Philimon N. and Gopalani, Sameer Vali and Gorman, Taren M. and Goulart, Alessandra C and Goulart, B{\'{a}}rbara Niegia Garcia and Grada, Ayman and Grams, Morgan E. and Grosso, Giuseppe and Gugnani, Harish Chander and Guo, Yuming and Gupta, Prakash C and Gupta, Rahul and Gupta, Rajeev and Gupta, Tanush and Gyawali, Bishal and Haagsma, Juanita A. and Hachinski, Vladimir and Hafezi-Nejad, Nima and {Haghparast Bidgoli}, Hassan and Hagos, Tekleberhan B. and Hailu, Gessessew Bugssa and Haj-Mirzaian, Arvin and Haj-Mirzaian, Arya and Hamadeh, Randah R. and Hamidi, Samer and Handal, Alexis J. and Hankey, Graeme J. and Hao, Yuantao and Harb, Hilda L. and Harikrishnan, Sivadasanpillai and Haro, Josep Maria and Hasan, Mehedi and Hassankhani, Hadi and Hassen, Hamid Yimam and Havmoeller, Rasmus and Hawley, Caitlin N. and Hay, Roderick J and Hay, Simon I. and Hedayatizadeh-Omran, Akbar and Heibati, Behzad and Hendrie, Delia and Henok, Andualem and Herteliu, Claudiu and Heydarpour, Sousan and Hibstu, Desalegn Tsegaw and Hoang, Huong Thanh and Hoek, Hans W. and Hoffman, Howard J. and Hole, Michael K. and {Homaie Rad}, Enayatollah and Hoogar, Praveen and Hosgood, H. Dean and Hosseini, Seyed Mostafa and Hosseinzadeh, Mehdi and Hostiuc, Mihaela and Hostiuc, Sorin and Hotez, Peter J. and Hoy, Damian G. and Hsairi, Mohamed and Htet, Aung Soe and Hu, Guoqing and Huang, John J. and Huynh, Chantal K. and Iburg, Kim Moesgaard and Ikeda, Chad Thomas and Ileanu, Bogdan and Ilesanmi, Olayinka Stephen and Iqbal, Usman and Irvani, Seyed Sina Naghibi and Irvine, Caleb Mackay Salpeter and Islam, Sheikh Mohammed Shariful and Islami, Farhad and Jacobsen, Kathryn H. and Jahangiry, Leila and Jahanmehr, Nader and Jain, Sudhir Kumar and Jakovljevic, Mihajlo and Javanbakht, Mehdi and Jayatilleke, Achala Upendra and Jeemon, Panniyammakal and Jha, Ravi Prakash and Jha, Vivekanand and Ji, John S. and Johnson, Catherine O. and Jonas, Jost B. and Jozwiak, Jacek Jerzy and Jungari, Suresh Banayya and J{\"{u}}risson, Mikk and Kabir, Zubair and Kadel, Rajendra and Kahsay, Amaha and Kalani, Rizwan and Kanchan, Tanuj and Karami, Manoochehr and {Karami Matin}, Behzad and Karch, Andr{\'{e}} and Karema, Corine and Karimi, Narges and Karimi, Seyed M. and Kasaeian, Amir and Kassa, Dessalegn H and Kassa, Getachew Mullu and Kassa, Tesfaye Dessale and Kassebaum, Nicholas J. and Katikireddi, Srinivasa Vittal and Kawakami, Norito and Karyani, Ali Kazemi and Keighobadi, Masoud Masoud and Keiyoro, Peter Njenga and Kemmer, Laura and Kemp, Grant Rodgers and Kengne, Andre Pascal and Keren, Andre and Khader, Yousef Saleh and Khafaei, Behzad and Khafaie, Morteza Abdullatif and Khajavi, Alireza and Khalil, Ibrahim A. and Khan, Ejaz Ahmad and Khan, Muhammad Shahzeb and Khan, Muhammad Ali and Khang, Young-Ho and Khazaei, Mohammad and Khoja, Abdullah T. and Khosravi, Ardeshir and Khosravi, Mohammad Hossein and Kiadaliri, Aliasghar A. and Kiirithio, Daniel N. and Kim, Cho-Il and Kim, Daniel and Kim, Pauline and Kim, Young-Eun and Kim, Yun Jin and Kimokoti, Ruth W. and Kinfu, Yohannes and Kisa, Adnan and Kissimova-Skarbek, Katarzyna and Kivim{\"{a}}ki, Mika and Knudsen, Ann Kristin Skrindo and Kocarnik, Jonathan M. and Kochhar, Sonali and Kokubo, Yoshihiro and Kolola, Tufa and Kopec, Jacek A. and Kosen, Soewarta and Kotsakis, Georgios A. and Koul, Parvaiz A. and Koyanagi, Ai and Kravchenko, Michael A. and Krishan, Kewal and Krohn, Kristopher J. and {Kuate Defo}, Barthelemy and {Kucuk Bicer}, Burcu and Kumar, G Anil and Kumar, Manasi and Kyu, Hmwe Hmwe and Lad, Deepesh P and Lad, Sheetal D. and Lafranconi, Alessandra and Lalloo, Ratilal and Lallukka, Tea and Lami, Faris Hasan and Lansingh, Van C. and Latifi, Arman and Lau, Kathryn Mei-Ming and Lazarus, Jeffrey V. and Leasher, Janet L. and Ledesma, Jorge R. and Lee, Paul H. and Leigh, James and Leung, Janni and Levi, Miriam and Lewycka, Sonia and Li, Shanshan and Li, Yichong and Liao, Yu and Liben, Misgan Legesse and Lim, Lee-Ling and Lim, Stephen S. and Liu, Shiwei and Lodha, Rakesh and Looker, Katharine J. and Lopez, Alan D. and Lorkowski, Stefan and Lotufo, Paulo A. and Low, Nicola and Lozano, Rafael and Lucas, Tim C D and Lucchesi, Lydia R. and Lunevicius, Raimundas and Lyons, Ronan A. and Ma, Stefan and Macarayan, Erlyn Rachelle King and Mackay, Mark T. and Madotto, Fabiana and {Magdy Abd El Razek}, Hassan and {Magdy Abd El Razek}, Muhammed and Maghavani, Dhaval P. and Mahotra, Narayan Bahadur and Mai, Hue Thi and Majdan, Marek and Majdzadeh, Reza and Majeed, Azeem and Malekzadeh, Reza and Malta, Deborah Carvalho and Mamun, Abdullah A. and Manda, Ana-Laura and Manguerra, Helena and Manhertz, Treh and Mansournia, Mohammad Ali and Mantovani, Lorenzo Giovanni and Mapoma, Chabila Christopher and Maravilla, Joemer C. and Marcenes, Wagner and Marks, Ashley and Martins-Melo, Francisco Rogerl{\^{a}}ndio and Martopullo, Ira and M{\"{a}}rz, Winfried and Marzan, Melvin B. and Mashamba-Thompson, Tivani Phosa and Massenburg, Benjamin Ballard and Mathur, Manu Raj and Matsushita, Kunihiro and Maulik, Pallab K. and Mazidi, Mohsen and McAlinden, Colm and McGrath, John J. and McKee, Martin and Mehndiratta, Man Mohan and Mehrotra, Ravi and Mehta, Kala M and Mehta, Varshil and Mejia-Rodriguez, Fabiola and Mekonen, Tesfa and Melese, Addisu and Melku, Mulugeta and Meltzer, Michele and Memiah, Peter T N and Memish, Ziad A. and Mendoza, Walter and Mengistu, Desalegn Tadese and Mengistu, Getnet and Mensah, George A. and Mereta, Seid Tiku and Meretoja, Atte and Meretoja, Tuomo J. and Mestrovic, Tomislav and Mezerji, Naser Mohammad Gholi and Miazgowski, Bartosz and Miazgowski, Tomasz and Millear, Anoushka I. and Miller, Ted R. and Miltz, Benjamin and Mini, G. K. and Mirarefin, Mojde and Mirrakhimov, Erkin M. and Misganaw, Awoke Temesgen and Mitchell, Philip B. and Mitiku, Habtamu and Moazen, Babak and Mohajer, Bahram and Mohammad, Karzan Abdulmuhsin and Mohammadifard, Noushin and Mohammadnia-Afrouzi, Mousa and Mohammed, Mohammed A and Mohammed, Shafiu and Mohebi, Farnam and Moitra, Modhurima and Mokdad, Ali H. and Molokhia, Mariam and Monasta, Lorenzo and Moodley, Yoshan and Moosazadeh, Mahmood and Moradi, Ghobad and Moradi-Lakeh, Maziar and Moradinazar, Mehdi and Moraga, Paula and Morawska, Lidia and {Moreno Vel{\'{a}}squez}, Ilais and Morgado-Da-Costa, Joana and Morrison, Shane Douglas and Moschos, Marilita M. and Mountjoy-Venning, W Cliff and Mousavi, Seyyed Meysam and Mruts, Kalayu Brhane and Muche, Achenef Asmamaw and Muchie, Kindie Fentahun and Mueller, Ulrich Otto and Muhammed, Oumer Sada and Mukhopadhyay, Satinath and Muller, Kate and Mumford, John Everett and Murhekar, Manoj and Musa, Jonah and Musa, Kamarul Imran and Mustafa, Ghulam and Nabhan, Ashraf F. and Nagata, Chie and Naghavi, Mohsen and Naheed, Aliya and Nahvijou, Azin and Naik, Gurudatta and Naik, Nitish and Najafi, Farid and Naldi, Luigi and Nam, Hae Sung and Nangia, Vinay and Nansseu, Jobert Richie and Nascimento, Bruno Ramos and Natarajan, Gopalakrishnan and Neamati, Nahid and Negoi, Ionut and Negoi, Ruxandra Irina and Neupane, Subas and Newton, Charles Richard James and Ngunjiri, Josephine W. and Nguyen, Anh Quynh and Nguyen, Ha Thu and Nguyen, Huong Lan Thi and Nguyen, Huong Thanh and Nguyen, Long Hoang and Nguyen, Minh and Nguyen, Nam Ba and Nguyen, Son Hoang and Nichols, Emma and Ningrum, Dina Nur Anggraini and Nixon, Molly R. and Nolutshungu, Nomonde and Nomura, Shuhei and Norheim, Ole F. and Noroozi, Mehdi and Norrving, Bo and Noubiap, Jean Jacques and Nouri, Hamid Reza and {Nourollahpour Shiadeh}, Malihe and Nowroozi, Mohammad Reza and Nsoesie, Elaine O. and Nyasulu, Peter S. and Odell, Christopher M. and Ofori-Asenso, Richard and Ogbo, Felix Akpojene and Oh, In-Hwan and Oladimeji, Olanrewaju and Olagunju, Andrew T and Olagunju, Tinuke O. and Olivares, Pedro R. and Olsen, Helen Elizabeth and Olusanya, Bolajoko Olubukunola and Ong, Kanyin L and Ong, Sok King and Oren, Eyal and Ortiz, Alberto and Ota, Erika and Otstavnov, Stanislav S. and {\O}verland, Simon and Owolabi, Mayowa Ojo and {P A}, Mahesh and Pacella, Rosana and Pakpour, Amir H. and Pana, Adrian and Panda-Jonas, Songhomitra and Parisi, Andrea and Park, Eun-Kee and Parry, Charles D H and Patel, Shanti and Pati, Sanghamitra and Patil, Snehal T. and Patle, Ajay and Patton, George C. and Paturi, Vishnupriya Rao and Paulson, Katherine R. and Pearce, Neil and Pereira, David M. and Perico, Norberto and Pesudovs, Konrad and Pham, Hai Quang and Phillips, Michael R. and Pigott, David M. and Pillay, Julian David and Piradov, Michael A. and Pirsaheb, Meghdad and Pishgar, Farhad and Plana-Ripoll, Oleguer and Plass, Dietrich and Polinder, Suzanne and Popova, Svetlana and Postma, Maarten J. and Pourshams, Akram and Poustchi, Hossein and Prabhakaran, Dorairaj and Prakash, Swayam and Prakash, V. and Purcell, Caroline A. and Purwar, Manorama B. and Qorbani, Mostafa and Quistberg, D. Alex and Radfar, Amir and Rafay, Anwar and Rafiei, Alireza and Rahim, Fakher and Rahimi, Kazem and Rahimi-Movaghar, Afarin and Rahimi-Movaghar, Vafa and Rahman, Mahfuzar and ur Rahman, Mohammad Hifz and Rahman, Muhammad Aziz and Rahman, Sajjad Ur and Rai, Rajesh Kumar and Rajati, Fatemeh and Ram, Usha and Ranjan, Prabhat and Ranta, Anna and Rao, Puja C. and Rawaf, David Laith and Rawaf, Salman and Reddy, K. Srinath and Reiner, Robert C. and Reinig, Nickolas and Reitsma, Marissa Bettay and Remuzzi, Giuseppe and Renzaho, Andre M N and Resnikoff, Serge and Rezaei, Satar and Rezai, Mohammad Sadegh and Ribeiro, Antonio Luiz P. and Roberts, Nicholas L S and Robinson, Stephen R. and Roever, Leonardo and Ronfani, Luca and Roshandel, Gholamreza and Rostami, Ali and Roth, Gregory A. and Roy, Ambuj and Rubagotti, Enrico and Sachdev, Perminder S. and Sadat, Nafis and Saddik, Basema and Sadeghi, Ehsan and {Saeedi Moghaddam}, Sahar and Safari, Hosein and Safari, Yahya and Safari-Faramani, Roya and Safdarian, Mahdi and Safi, Sare and Safiri, Saeid and Sagar, Rajesh and Sahebkar, Amirhossein and Sahraian, Mohammad Ali and Sajadi, Haniye Sadat and Salam, Nasir and Salama, Joseph S. and Salamati, Payman and Saleem, Komal and Saleem, Zikria and Salimi, Yahya and Salomon, Joshua A. and Salvi, Sundeep Santosh and Salz, Inbal and Samy, Abdallah M. and Sanabria, Juan and Sang, Yingying and Santomauro, Damian Francesco and Santos, Itamar S and Santos, Jo{\~{a}}o Vasco and {Santric Milicevic}, Milena M. and {Sao Jose}, Bruno Piassi and Sardana, Mayank and Sarker, Abdur Razzaque and Sarrafzadegan, Nizal and Sartorius, Benn and Sarvi, Shahabeddin and Sathian, Brijesh and Satpathy, Maheswar and Sawant, Arundhati R. and Sawhney, Monika and Saxena, Sonia and Saylan, Mete and Schaeffner, Elke and Schmidt, Maria In{\^{e}}s and Schneider, Ione J C and Sch{\"{o}}ttker, Ben and Schwebel, David C. and Schwendicke, Falk and Scott, James G. and Sekerija, Mario and Sepanlou, Sadaf G. and Serv{\'{a}}n-Mori, Edson and Seyedmousavi, Seyedmojtaba and Shabaninejad, Hosein and Shafieesabet, Azadeh and Shahbazi, Mehdi and Shaheen, Amira A. and Shaikh, Masood Ali and Shams-Beyranvand, Mehran and Shamsi, Mohammadbagher and Shamsizadeh, Morteza and Sharafi, Heidar and Sharafi, Kiomars and Sharif, Mehdi and Sharif-Alhoseini, Mahdi and Sharma, Meenakshi and Sharma, Rajesh and She, Jun and Sheikh, Aziz and Shi, Peilin and Shibuya, Kenji and Shigematsu, Mika and Shiri, Rahman and Shirkoohi, Reza and Shishani, Kawkab and Shiue, Ivy and Shokraneh, Farhad and Shoman, Haitham and Shrime, Mark G. and Si, Si and Siabani, Soraya and Siddiqi, Tariq J. and Sigfusdottir, Inga Dora and Sigurvinsdottir, Rannveig and Silva, Jo{\~{a}}o Pedro and Silveira, Dayane Gabriele Alves and Singam, Narayana Sarma Venkata and Singh, Jasvinder A and Singh, Narinder Pal and Singh, Virendra and Sinha, Dhirendra Narain and Skiadaresi, Eirini and Slepak, Erica Leigh N. and Sliwa, Karen and Smith, David L and Smith, Mari and {Soares Filho}, Adauto Martins and Sobaih, Badr Hasan and Sobhani, Soheila and Sobngwi, Eug{\`{e}}ne and Soneji, Samir S. and Soofi, Moslem and Soosaraei, Masoud and Sorensen, Reed J D and Soriano, Joan B. and Soyiri, Ireneous N. and Sposato, Luciano A. and Sreeramareddy, Chandrashekhar T. and Srinivasan, Vinay and Stanaway, Jeffrey D. and Stein, Dan J. and Steiner, Caitlyn and Steiner, Timothy J. and Stokes, Mark A. and Stovner, Lars Jacob and Subart, Michelle L. and Sudaryanto, Agus and Sufiyan, Mu'awiyyah Babale and Sunguya, Bruno F. and Sur, Patrick John and Sutradhar, Ipsita and Sykes, Bryan L. and Sylte, Dillon O. and Tabar{\'{e}}s-Seisdedos, Rafael and Tadakamadla, Santosh Kumar and Tadesse, Birkneh Tilahun and Tandon, Nikhil and Tassew, Segen Gebremeskel and Tavakkoli, Mohammad and Taveira, Nuno and Taylor, Hugh R. and Tehrani-Banihashemi, Arash and Tekalign, Tigist Gashaw and Tekelemedhin, Shishay Wahdey and Tekle, Merhawi Gebremedhin and Temesgen, Habtamu and Temsah, Mohamad-Hani and Temsah, Omar and Terkawi, Abdullah Sulieman and Teweldemedhin, Mebrahtu and Thankappan, Kavumpurathu Raman and Thomas, Nihal and Tilahun, Binyam and To, Quyen G. and Tonelli, Marcello and Topor-Madry, Roman and Topouzis, Fotis and Torre, Anna E. and Tortajada-Girb{\'{e}}s, Miguel and Touvier, Mathilde and Tovani-Palone, Marcos Roberto and Towbin, Jeffrey A. and Tran, Bach Xuan and Tran, Khanh Bao and Troeger, Christopher E. and Truelsen, Thomas Clement and Tsilimbaris, Miltiadis K. and Tsoi, Derrick and {Tudor Car}, Lorainne and Tuzcu, E. Murat and Ukwaja, Kingsley N. and Ullah, Irfan and Undurraga, Eduardo A. and Unutzer, Jurgen and Updike, Rachel L. and Usman, Muhammad Shariq and Uthman, Olalekan A. and Vaduganathan, Muthiah and Vaezi, Afsane and Valdez, Pascual R. and Varughese, Santosh and Vasankari, Tommi Juhani and Venketasubramanian, Narayanaswamy and Villafaina, Santos and Violante, Francesco S. and Vladimirov, Sergey Konstantinovitch and Vlassov, Vasily and Vollset, Stein Emil and Vosoughi, Kia and Vujcic, Isidora S. and Wagnew, Fasil Shiferaw and Waheed, Yasir and Waller, Stephen G. and Wang, Yafeng and Wang, Yuan-Pang and Weiderpass, Elisabete and Weintraub, Robert G. and Weiss, Daniel J. and Weldegebreal, Fitsum and Weldegwergs, Kidu Gidey and Werdecker, Andrea and West, T. Eoin and Whiteford, Harvey A. and Widecka, Justyna and Wijeratne, Tissa and Wilner, Lauren B. and Wilson, Shadrach and Winkler, Andrea Sylvia and Wiyeh, Alison B. and Wiysonge, Charles Shey and Wolfe, Charles D A and Woolf, Anthony D. and Wu, Shouling and Wu, Yun-Chun and Wyper, Grant M A and Xavier, Denis and Xu, Gelin and Yadgir, Simon and Yadollahpour, Ali and {Yahyazadeh Jabbari}, Seyed Hossein and Yamada, Tomohide and Yan, Lijing L. and Yano, Yuichiro and Yaseri, Mehdi and Yasin, Yasin Jemal and Yeshaneh, Alex and Yimer, Ebrahim M. and Yip, Paul and Yisma, Engida and Yonemoto, Naohiro and Yoon, Seok-Jun and Yotebieng, Marcel and Younis, Mustafa Z. and Yousefifard, Mahmoud and Yu, Chuanhua and Zadnik, Vesna and Zaidi, Zoubida and Zaman, Sojib Bin and Zamani, Mohammad and Zare, Zohreh and Zeleke, Ayalew Jejaw and Zenebe, Zerihun Menlkalew and Zhang, Kai and Zhao, Zheng and Zhou, Maigeng and Zodpey, Sanjay and Zucker, Inbar and Vos, Theo and Murray, Christopher J L},
doi = {10.1016/S0140-6736(18)32279-7},
issn = {01406736},
journal = {The Lancet},
month = {nov},
number = {10159},
pages = {1789--1858},
pmid = {30496104},
title = {{Global, regional, and national incidence, prevalence, and years lived with disability for 354 diseases and injuries for 195 countries and territories, 1990–2017: a systematic analysis for the Global Burden of Disease Study 2017}},
url = {https://linkinghub.elsevier.com/retrieve/pii/S0140673618322797},
volume = {392},
year = {2018}
}

@article{Stade2024,
author = {Stade, Elizabeth C. and Stirman, Shannon Wiltsey and Ungar, Lyle H. and Boland, Cody L. and Schwartz, H. Andrew and Yaden, David B. and Sedoc, Jo{\~{a}}o and DeRubeis, Robert J. and Willer, Robb and Eichstaedt, Johannes C.},
doi = {10.1038/s44184-024-00056-z},
issn = {2731-4251},
journal = {npj Mental Health Research},
month = {apr},
number = {1},
pages = {12},
title = {{Large language models could change the future of behavioral healthcare: a proposal for responsible development and evaluation}},
url = {https://www.nature.com/articles/s44184-024-00056-z},
volume = {3},
year = {2024}
}

@inproceedings{chaszczewicz-etal-2024-multi,
    title = "Multi-Level Feedback Generation with Large Language Models for Empowering Novice Peer Counselors",
    author = "Chaszczewicz, Alicja  and
      Shah, Raj  and
      Louie, Ryan  and
      Arnow, Bruce  and
      Kraut, Robert  and
      Yang, Diyi",
    editor = "Ku, Lun-Wei  and
      Martins, Andre  and
      Srikumar, Vivek",
    booktitle = "Proceedings of the 62nd Annual Meeting of the Association for Computational Linguistics (Volume 1: Long Papers)",
    month = aug,
    year = "2024",
    address = "Bangkok, Thailand",
    publisher = "Association for Computational Linguistics",
    url = "https://aclanthology.org/2024.acl-long.227",
    doi = "10.18653/v1/2024.acl-long.227",
    pages = "4130--4161",
}

@inproceedings{10.1145/3613904.3642761,
author = {Sharma, Ashish and Rushton, Kevin and Lin, Inna Wanyin and Nguyen, Theresa and Althoff, Tim},
title = {Facilitating Self-Guided Mental Health Interventions Through Human-Language Model Interaction: A Case Study of Cognitive Restructuring},
year = {2024},
isbn = {9798400703300},
publisher = {Association for Computing Machinery},
address = {New York, NY, USA},
url = {https://doi.org/10.1145/3613904.3642761},
doi = {10.1145/3613904.3642761},
booktitle = {Proceedings of the 2024 CHI Conference on Human Factors in Computing Systems},
articleno = {700},
numpages = {29},
location = {Honolulu, HI, USA},
series = {CHI '24}
}

@inproceedings{10.1145/3613904.3642937,
author = {Kim, Taewan and Bae, Seolyeong and Kim, Hyun Ah and Lee, Su-Woo and Hong, Hwajung and Yang, Chanmo and Kim, Young-Ho},
title = {MindfulDiary: Harnessing Large Language Model to Support Psychiatric Patients' Journaling},
year = {2024},
isbn = {9798400703300},
publisher = {Association for Computing Machinery},
address = {New York, NY, USA},
url = {https://doi.org/10.1145/3613904.3642937},
doi = {10.1145/3613904.3642937},
booktitle = {Proceedings of the 2024 CHI Conference on Human Factors in Computing Systems},
articleno = {701},
numpages = {20},
location = {Honolulu, HI, USA},
series = {CHI '24}
}

@article{schueller2021understanding,
  title={Understanding people’s use of and perspectives on mood-tracking apps: interview study},
  author={Schueller, Stephen M and Neary, Martha and Lai, Jocelyn and Epstein, Daniel A},
  journal={JMIR mental health},
  volume={8},
  number={8},
  pages={e29368},
  year={2021},
  publisher={JMIR Publications Toronto, Canada}
}

@article{meyerhoff2024small,
  title={Small Steps over time: A longitudinal usability test of an automated interactive text messaging intervention to support self-management of depression and anxiety symptoms},
  author={Meyerhoff, Jonah and Beltzer, Miranda and Popowski, Sarah and Karr, Chris J and Nguyen, Theresa and Williams, Joseph J and Krause, Charles J and Kumar, Harsh and Bhattacharjee, Ananya and Mohr, David C and others},
  journal={Journal of Affective Disorders},
  volume={345},
  pages={122--130},
  year={2024},
  publisher={Elsevier}
}

@inproceedings{
mireshghallah2024can,
title={Can {LLM}s Keep a Secret? Testing  Privacy  Implications of Language Models  via Contextual Integrity Theory},
author={Niloofar Mireshghallah and Hyunwoo Kim and Xuhui Zhou and Yulia Tsvetkov and Maarten Sap and Reza Shokri and Yejin Choi},
booktitle={The Twelfth International Conference on Learning Representations},
year={2024},
url={https://openreview.net/forum?id=gmg7t8b4s0}
}

@inproceedings{carlini2019secret,
  title={The secret sharer: Evaluating and testing unintended memorization in neural networks},
  author={Carlini, Nicholas and Liu, Chang and Erlingsson, {\'U}lfar and Kos, Jernej and Song, Dawn},
  booktitle={28th USENIX security symposium (USENIX security 19)},
  pages={267--284},
  year={2019}
}

@article{Huang2023ASO,
  title={A Survey on Hallucination in Large Language Models: Principles, Taxonomy, Challenges, and Open Questions},
  author={Lei Huang and Weijiang Yu and Weitao Ma and Weihong Zhong and Zhangyin Feng and Haotian Wang and Qianglong Chen and Weihua Peng and Xiaocheng Feng and Bing Qin and Ting Liu},
  journal={ArXiv},
  year={2023},
  volume={abs/2311.05232},
  url={https://api.semanticscholar.org/CorpusID:265067168}
}

@misc{gabriel2024airelatetestinglarge,
      title={Can AI Relate: Testing Large Language Model Response for Mental Health Support}, 
      author={Saadia Gabriel and Isha Puri and Xuhai Xu and Matteo Malgaroli and Marzyeh Ghassemi},
      year={2024},
      eprint={2405.12021},
      archivePrefix={arXiv},
      primaryClass={cs.CL},
      url={https://arxiv.org/abs/2405.12021}, 
}

@article{omiye2023large,
  title={Large language models propagate race-based medicine},
  author={Omiye, Jesutofunmi A and Lester, Jenna C and Spichak, Simon and Rotemberg, Veronica and Daneshjou, Roxana},
  journal={NPJ Digital Medicine},
  volume={6},
  number={1},
  pages={195},
  year={2023},
  publisher={Nature Publishing Group UK London}
}

@article{gallegos2024bias,
  title={Bias and fairness in large language models: A survey},
  author={Gallegos, Isabel O and Rossi, Ryan A and Barrow, Joe and Tanjim, Md Mehrab and Kim, Sungchul and Dernoncourt, Franck and Yu, Tong and Zhang, Ruiyi and Ahmed, Nesreen K},
  journal={Computational Linguistics},
  pages={1--79},
  year={2024},
  publisher={MIT Press 255 Main Street, 9th Floor, Cambridge, Massachusetts 02142, USA~…}
}

@inproceedings{
sharma2024towards,
title={Towards Understanding Sycophancy in Language Models},
author={Mrinank Sharma and Meg Tong and Tomasz Korbak and David Duvenaud and Amanda Askell and Samuel R. Bowman and Esin DURMUS and Zac Hatfield-Dodds and Scott R Johnston and Shauna M Kravec and Timothy Maxwell and Sam McCandlish and Kamal Ndousse and Oliver Rausch and Nicholas Schiefer and Da Yan and Miranda Zhang and Ethan Perez},
booktitle={The Twelfth International Conference on Learning Representations},
year={2024},
url={https://openreview.net/forum?id=tvhaxkMKAn}
}

@inproceedings{ma2023understanding,
  title={Understanding the benefits and challenges of using large language model-based conversational agents for mental well-being support},
  author={Ma, Zilin and Mei, Yiyang and Su, Zhaoyuan},
  booktitle={AMIA Annual Symposium Proceedings},
  volume={2023},
  pages={1105},
  year={2023},
  organization={American Medical Informatics Association}
}

@article{de2024chatbots,
  title={Chatbots and mental health: Insights into the safety of generative AI},
  author={De Freitas, Julian and U{\u{g}}uralp, Ahmet Kaan and O{\u{g}}uz-U{\u{g}}uralp, Zeliha and Puntoni, Stefano},
  journal={Journal of Consumer Psychology},
  volume={34},
  number={3},
  pages={481--491},
  year={2024},
  publisher={Wiley Online Library}
}

@article{heston2023safety,
  title={Safety of large language models in addressing depression},
  author={Heston, Thomas F},
  journal={Cureus},
  volume={15},
  number={12},
  year={2023},
  publisher={Cureus Inc.}
}

% \appendix

% \section{Why tokens and not words}
% \label{app:tokens}

% %BS: do not forget about this one :)

\end{document}